\documentclass[twocolumn,aps,prc,longbibliography,superscriptaddress,nofootinbib,showpacs,floatfix]{revtex4-1}

\usepackage{graphicx}	
\usepackage{array}

\usepackage{xspace}	
\usepackage{amsmath}

\newcommand{\pt}{\mbox{$p_T$}\xspace}

\newcommand{\sqsn}{\mbox{$\sqrt{s_{_{NN}}}$}\xspace}

\newcommand{\mt}{\mbox{$m_T$}\xspace}
\newcommand{\kt}{\mbox{$k_T$}\xspace}
\newcommand{\Rs}{\mbox{$R_{\rm s}$}\xspace}
\newcommand{\Ro}{\mbox{$R_{\rm o}$}\xspace}
\newcommand{\Rl}{\mbox{$R_{\rm l}$}\xspace}
\newcommand{\qs}{\mbox{$q_{\rm s}$}\xspace}
\newcommand{\qo}{\mbox{$q_{\rm o}$}\xspace}
\newcommand{\ql}{\mbox{$q_{\rm l}$}\xspace}
\newcommand{\Ros}{\mbox{$R_{\rm os}$}\xspace}

\newcolumntype{X}{>{\centering\arraybackslash}p{10em}} 

\begin{document}

\title{Systematic study of charged-pion and kaon femtoscopy in Au$+$Au 
collisions at $\sqrt{s_{_{NN}}}$=200 GeV}

\newcommand{\abilene}{Abilene Christian University, Abilene, Texas 79699, USA}
\newcommand{\augie}{Department of Physics, Augustana College, Sioux Falls, South Dakota 57197, USA}
\newcommand{\banaras}{Department of Physics, Banaras Hindu University, Varanasi 221005, India}
\newcommand{\barc}{Bhabha Atomic Research Centre, Bombay 400 085, India}
\newcommand{\baruch}{Baruch College, City University of New York, New York, New York, 10010 USA}
\newcommand{\bnlcoll}{Collider-Accelerator Department, Brookhaven National Laboratory, Upton, New York 11973-5000, USA}
\newcommand{\bnlphys}{Physics Department, Brookhaven National Laboratory, Upton, New York 11973-5000, USA}
\newcommand{\caucr}{University of California-Riverside, Riverside, California 92521, USA}
\newcommand{\charlesczech}{Charles University, Ovocn\'{y} trh 5, Praha 1, 116 36, Prague, Czech Republic}
\newcommand{\chonbuk}{Chonbuk National University, Jeonju, 561-756, Korea}
\newcommand{\ciae}{Science and Technology on Nuclear Data Laboratory, China Institute of Atomic Energy, Beijing 102413, P.~R.~China}
\newcommand{\cns}{Center for Nuclear Study, Graduate School of Science, University of Tokyo, 7-3-1 Hongo, Bunkyo, Tokyo 113-0033, Japan}
\newcommand{\colorado}{University of Colorado, Boulder, Colorado 80309, USA}
\newcommand{\columbia}{Columbia University, New York, New York 10027 and Nevis Laboratories, Irvington, New York 10533, USA}
\newcommand{\czechtech}{Czech Technical University, Zikova 4, 166 36 Prague 6, Czech Republic}
\newcommand{\dapnia}{Dapnia, CEA Saclay, F-91191, Gif-sur-Yvette, France}
\newcommand{\debrecen}{Debrecen University, H-4010 Debrecen, Egyetem t{\'e}r 1, Hungary}
\newcommand{\elte}{ELTE, E{\"o}tv{\"o}s Lor{\'a}nd University, H-1117 Budapest, P{\'a}zm{\'a}ny P.~s.~1/A, Hungary}
\newcommand{\ewha}{Ewha Womans University, Seoul 120-750, Korea}
\newcommand{\fit}{Florida Institute of Technology, Melbourne, Florida 32901, USA}
\newcommand{\fsu}{Florida State University, Tallahassee, Florida 32306, USA}
\newcommand{\gsu}{Georgia State University, Atlanta, Georgia 30303, USA}
\newcommand{\hiroshima}{Hiroshima University, Kagamiyama, Higashi-Hiroshima 739-8526, Japan}
\newcommand{\howard}{Department of Physics and Astronomy, Howard University, Washington, DC 20059, USA}
\newcommand{\ihepprot}{IHEP Protvino, State Research Center of Russian Federation, Institute for High Energy Physics, Protvino, 142281, Russia}
\newcommand{\illuiuc}{University of Illinois at Urbana-Champaign, Urbana, Illinois 61801, USA}
\newcommand{\inrras}{Institute for Nuclear Research of the Russian Academy of Sciences, prospekt 60-letiya Oktyabrya 7a, Moscow 117312, Russia}
\newcommand{\instpasczech}{Institute of Physics, Academy of Sciences of the Czech Republic, Na Slovance 2, 182 21 Prague 8, Czech Republic}
\newcommand{\isu}{Iowa State University, Ames, Iowa 50011, USA}
\newcommand{\jaea}{Advanced Science Research Center, Japan Atomic Energy Agency, 2-4 Shirakata Shirane, Tokai-mura, Naka-gun, Ibaraki-ken 319-1195, Japan}
\newcommand{\jinrdubna}{Joint Institute for Nuclear Research, 141980 Dubna, Moscow Region, Russia}
\newcommand{\jyvaskyla}{Helsinki Institute of Physics and University of Jyv{\"a}skyl{\"a}, P.O.Box 35, FI-40014 Jyv{\"a}skyl{\"a}, Finland}
\newcommand{\kek}{KEK, High Energy Accelerator Research Organization, Tsukuba, Ibaraki 305-0801, Japan}
\newcommand{\korea}{Korea University, Seoul, 136-701, Korea}
\newcommand{\kurchatov}{Russian Research Center ``Kurchatov Institute", Moscow, 123098 Russia}
\newcommand{\kyoto}{Kyoto University, Kyoto 606-8502, Japan}
\newcommand{\labllr}{Laboratoire Leprince-Ringuet, Ecole Polytechnique, CNRS-IN2P3, Route de Saclay, F-91128, Palaiseau, France}
\newcommand{\lahorelums}{Physics Department, Lahore University of Management Sciences, Lahore 54792, Pakistan}
\newcommand{\lawllnl}{Lawrence Livermore National Laboratory, Livermore, California 94550, USA}
\newcommand{\losalamos}{Los Alamos National Laboratory, Los Alamos, New Mexico 87545, USA}
\newcommand{\lpc}{LPC, Universit{\'e} Blaise Pascal, CNRS-IN2P3, Clermont-Fd, 63177 Aubiere Cedex, France}
\newcommand{\lund}{Department of Physics, Lund University, Box 118, SE-221 00 Lund, Sweden}
\newcommand{\maryland}{University of Maryland, College Park, Maryland 20742, USA}
\newcommand{\mass}{Department of Physics, University of Massachusetts, Amherst, Massachusetts 01003-9337, USA}
\newcommand{\michigan}{Department of Physics, University of Michigan, Ann Arbor, Michigan 48109-1040, USA}
\newcommand{\muenster}{Institut f\"ur Kernphysik, University of Muenster, D-48149 Muenster, Germany}
\newcommand{\muhlenberg}{Muhlenberg College, Allentown, Pennsylvania 18104-5586, USA}
\newcommand{\myongji}{Myongji University, Yongin, Kyonggido 449-728, Korea}
\newcommand{\nagasaki}{Nagasaki Institute of Applied Science, Nagasaki-shi, Nagasaki 851-0193, Japan}
\newcommand{\natmephi}{National Research Nuclear University, MEPhI, Moscow Engineering Physics Institute, Moscow, 115409, Russia}
\newcommand{\newmex}{University of New Mexico, Albuquerque, New Mexico 87131, USA}
\newcommand{\nmsu}{New Mexico State University, Las Cruces, New Mexico 88003, USA}
\newcommand{\ohio}{Department of Physics and Astronomy, Ohio University, Athens, Ohio 45701, USA}
\newcommand{\ornl}{Oak Ridge National Laboratory, Oak Ridge, Tennessee 37831, USA}
\newcommand{\orsay}{IPN-Orsay, Universite Paris Sud, CNRS-IN2P3, BP1, F-91406, Orsay, France}
\newcommand{\peking}{Peking University, Beijing 100871, P.~R.~China}
\newcommand{\pnpi}{PNPI, Petersburg Nuclear Physics Institute, Gatchina, Leningrad region, 188300, Russia}
\newcommand{\riken}{RIKEN Nishina Center for Accelerator-Based Science, Wako, Saitama 351-0198, Japan}
\newcommand{\rikjrbrc}{RIKEN BNL Research Center, Brookhaven National Laboratory, Upton, New York 11973-5000, USA}
\newcommand{\rikkyo}{Physics Department, Rikkyo University, 3-34-1 Nishi-Ikebukuro, Toshima, Tokyo 171-8501, Japan}
\newcommand{\saispbstu}{Saint Petersburg State Polytechnic University, St.~Petersburg, 195251 Russia}
\newcommand{\saopaulo}{Universidade de S{\~a}o Paulo, Instituto de F\'{\i}sica, Caixa Postal 66318, S{\~a}o Paulo CEP05315-970, Brazil}
\newcommand{\seoulnat}{Department of Physics and Astronomy, Seoul National University, Seoul 151-742, Korea}
\newcommand{\stonybrkc}{Chemistry Department, Stony Brook University, SUNY, Stony Brook, New York 11794-3400, USA}
\newcommand{\stonycrkp}{Department of Physics and Astronomy, Stony Brook University, SUNY, Stony Brook, New York 11794-3800, USA}
\newcommand{\tenn}{University of Tennessee, Knoxville, Tennessee 37996, USA}
\newcommand{\titech}{Department of Physics, Tokyo Institute of Technology, Oh-okayama, Meguro, Tokyo 152-8551, Japan}
\newcommand{\tsukuba}{Institute of Physics, University of Tsukuba, Tsukuba, Ibaraki 305, Japan}
\newcommand{\vandy}{Vanderbilt University, Nashville, Tennessee 37235, USA}
\newcommand{\waseda}{Waseda University, Advanced Research Institute for Science and Engineering, 17  Kikui-cho, Shinjuku-ku, Tokyo 162-0044, Japan}
\newcommand{\weizmann}{Weizmann Institute, Rehovot 76100, Israel}
\newcommand{\wigner}{Institute for Particle and Nuclear Physics, Wigner Research Centre for Physics, Hungarian Academy of Sciences (Wigner RCP, RMKI) H-1525 Budapest 114, POBox 49, Budapest, Hungary}
\newcommand{\yonsei}{Yonsei University, IPAP, Seoul 120-749, Korea}
\newcommand{\zagreb}{University of Zagreb, Faculty of Science, Department of Physics, Bijeni\v{c}ka 32, HR-10002 Zagreb, Croatia}
\affiliation{\abilene}
\affiliation{\augie}
\affiliation{\banaras}
\affiliation{\barc}
\affiliation{\baruch}
\affiliation{\bnlcoll}
\affiliation{\bnlphys}
\affiliation{\caucr}
\affiliation{\charlesczech}
\affiliation{\chonbuk}
\affiliation{\ciae}
\affiliation{\cns}
\affiliation{\colorado}
\affiliation{\columbia}
\affiliation{\czechtech}
\affiliation{\dapnia}
\affiliation{\debrecen}
\affiliation{\elte}
\affiliation{\ewha}
\affiliation{\fit}
\affiliation{\fsu}
\affiliation{\gsu}
\affiliation{\hiroshima}
\affiliation{\howard}
\affiliation{\ihepprot}
\affiliation{\illuiuc}
\affiliation{\inrras}
\affiliation{\instpasczech}
\affiliation{\isu}
\affiliation{\jaea}
\affiliation{\jinrdubna}
\affiliation{\jyvaskyla}
\affiliation{\kek}
\affiliation{\korea}
\affiliation{\kurchatov}
\affiliation{\kyoto}
\affiliation{\labllr}
\affiliation{\lahorelums}
\affiliation{\lawllnl}
\affiliation{\losalamos}
\affiliation{\lpc}
\affiliation{\lund}
\affiliation{\maryland}
\affiliation{\mass}
\affiliation{\michigan}
\affiliation{\muenster}
\affiliation{\muhlenberg}
\affiliation{\myongji}
\affiliation{\nagasaki}
\affiliation{\natmephi}
\affiliation{\newmex}
\affiliation{\nmsu}
\affiliation{\ohio}
\affiliation{\ornl}
\affiliation{\orsay}
\affiliation{\peking}
\affiliation{\pnpi}
\affiliation{\riken}
\affiliation{\rikjrbrc}
\affiliation{\rikkyo}
\affiliation{\saispbstu}
\affiliation{\saopaulo}
\affiliation{\seoulnat}
\affiliation{\stonybrkc}
\affiliation{\stonycrkp}
\affiliation{\tenn}
\affiliation{\titech}
\affiliation{\tsukuba}
\affiliation{\vandy}
\affiliation{\waseda}
\affiliation{\weizmann}
\affiliation{\wigner}
\affiliation{\yonsei}
\affiliation{\zagreb}
\author{A.~Adare} \affiliation{\colorado} 
\author{S.~Afanasiev} \affiliation{\jinrdubna} 
\author{C.~Aidala} \affiliation{\mass} \affiliation{\michigan} 
\author{N.N.~Ajitanand} \affiliation{\stonybrkc} 
\author{Y.~Akiba} \affiliation{\riken} \affiliation{\rikjrbrc} 
\author{H.~Al-Bataineh} \affiliation{\nmsu} 
\author{J.~Alexander} \affiliation{\stonybrkc} 
\author{M.~Alfred} \affiliation{\howard} 
\author{K.~Aoki} \affiliation{\kek} \affiliation{\kyoto} \affiliation{\riken} 
\author{N.~Apadula} \affiliation{\isu} \affiliation{\stonycrkp} 
% ...
\collaboration{PHENIX Collaboration} \noaffiliation

\date{\today}

\begin{abstract}

We present a systematic study of charged pion and kaon interferometry 
in Au$+$Au collisions at $\sqrt{s_{_{NN}}}$=200 GeV. The kaon mean 
source radii are found to be larger than pion radii in the outward 
and longitudinal directions for the same transverse mass; this 
difference increases for more central collisions.  The 
azimuthal-angle dependence of the radii was measured with respect to 
the second-order event plane and similar oscillations of the source 
radii were found for pions and kaons.  Hydrodynamic models 
qualitatively describe the similar oscillations of the mean source 
radii for pions and kaons, but they do not fully describe the 
transverse-mass dependence of the oscillations.

\end{abstract}

	
\maketitle


\section{Introduction\label{sec:intro}}

Measurements of the quark-gluon plasma (QGP), produced in nucleus-nucleus 
collisions at the Relativistic Heavy Ion Collider 
(RHIC)~\cite{wh_phenix,wh_star,wh_phobos,wh_brahms} and the Large Hadron 
Collider (LHC)~\cite{Raa_alice,jet_cms,dijet_atlas}, showed that the QGP 
exhibits rapid hydrodynamic expansion, followed by hadronization, which 
results in the emission of many particles. The time of last scattering 
among hadrons is referred to as kinetic freeze out. To understand 
the dynamics and properties of the QGP, it is important to understand the 
full system evolution and how it is constrained by the measurements of the 
space-time distribution at kinetic freeze out.

The quantum statistical interferometry of identical particles, also known 
as Hanbury Brown Twiss (HBT) interferometry or femtoscopy, is a powerful 
tool to measure the spatial and temporal scales of systems created in 
nucleus-nucleus collisions~\cite{bevalac,femtoscopy}. This technique was 
first developed to measure the angular diameter of stars through intensity 
interferometry of radio waves~\cite{hbt_org}. It has also been applied to 
nuclear and particle physics~\cite{GGLP}. In nucleus-nucleus collisions, 
the interferometry using emitted hadrons measures the spatial extent of 
the particle-emitting source at the time of kinetic freeze out.

Despite the successful description of various observables at RHIC by the 
hydrodynamic models~\cite{wh_phenix,wh_star}, there was significant 
discrepancy between HBT data and theoretical models~\cite{pihbt,wh_star}. 
Recent theoretical development has improved the agreement by including 
realistic physics conditions such as stiffer equation of state and a 
viscosity of the created matter~\cite{ResolvePuzzle}.

Charged pions are often used for the interferometry analysis because of 
their abundant production, but recently acquired large data sets by RHIC 
and LHC experiments allow study of the particle-species 
dependence~\cite{ppg82,k3Dsource,piKpHBT_ALICE}.  Kaon interferometry is of 
particular interest, because the contribution from resonance decays is 
reduced compared to pions~\cite{resonance1,resonance2}, thereby providing 
a more direct view of the particle-emitting source.  
PHENIX at RHIC published an analysis of 
one-dimensional source imaging for charged kaons 
~\cite{ppg82}.  STAR at RHIC has recently 
published three-dimensional source imaging~\cite{k3Dsource}, where 
charged kaons lack the nonGaussian tail in the source function observed 
in the pion sample.  This result may be caused by the reduced contribution 
from long-lived resonances as well as a different time dependence due to 
a shorter rescattering phase. Further systematic studies using different 
particle species are needed to better constrain the space-time evolution 
and freeze-out distributions of the created medium.

The HBT measurement is also sensitive to the initial spatial anisotropy 
and the subsequent evolution of the created matter. Due to the strong 
collective expansion, one might expect the eccentricity of the source 
shape in the initial state to be reduced at freeze out and possibly to be 
reversed if the collective expansion is stronger in the direction of the 
reaction plane, or if the expansion time is sufficiently long. To probe 
the spatial source anisotropy at freeze out, HBT measurements with respect 
to the event planes have been performed using two-pion 
correlation~\cite{AzHBTatAGS,AzHBTatSPS,star_azhbt,ppg149}. Large 
oscillations of the pion source radii relative to the second-order event 
plane were observed, which indicates that the pion source at freeze out is 
elongated in the direction perpendicular to the event plane even after the 
collective expansion.

In this paper, we present azimuthal-integrated and azimuthal-dependent 
source radii for charged pions and kaons in Au$+$Au collisions at $\sqsn$= 
200 GeV. Results are compared with the hydrodynamic models for both 
particle species.

 
\section{Experiment\label{sec:exp}}

\begin{figure}[htb]
\includegraphics[width=1.0\linewidth]{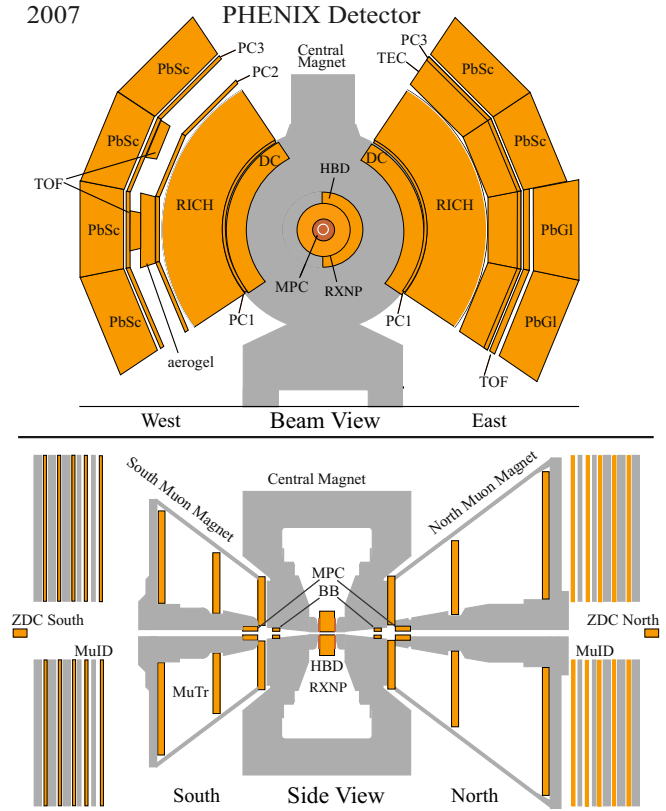}
\caption{(Color online) 
The layout of PHENIX detectors in the 2007 run configuration. The top 
figure shows the central arm detectors viewed along the beam axis. The 
bottom figure shows the side view of the global detectors and muon arm 
detectors.}
\label{fig1}
\end{figure}

The PHENIX experiment~\cite{PHENIXDoverview} is designed to measure 
particles produced in nucleus-nucleus collisions with good momentum 
resolution, including photons, electrons, muons, and hadrons, to 
study properties of the QGP.  The PHENIX detectors are comprised of magnet 
systems and detectors for particle tracking and identification, 
event timing, plus vertex position and centrality determination. The
particle tracking and identification detectors are arranged into central 
and forward (muon) arms.  Figure~\ref{fig1} shows the layout 
of the PHENIX detector during the 2007 running period.

Global detectors characterize the global event characteristics in heavy 
ion collisions. The beam-beam counters (BBC)~\cite{PHENIXDinner} measure 
the collision time, and the position of the collision vertex along the 
beam axis, as well as the collision centrality. The BBC comprises two 
identical sets of counters located $\pm$144 cm from the nominal collision 
point and surrounds the beam pipe covering the pseudorapidity range of 
$3.0<|\eta|<3.9$.  Each BBC has 64 modules of \v{C}erenkov radiators, and 
measures the number of charged particles in its acceptance. The 
zero-degree calorimeters are located 18 m from the nominal collision 
point and measure the energy of spectator neutrons. The reaction-plane 
detector (RXNP)~\cite{RXNP} was installed prior to the 2007 RHIC run to 
measure the event-plane angle in heavy ion collisions. The RXNP 
comprises two sets of 24 scintillators on both north and south sides, 
and is located $\pm$39 cm from the vertex position. The scintillators are 
arranged around the beam pipe in two concentric rings of 12 segments in 
the azimuthal direction. The outer and inner rings cover pseudorapidity 
ranges of $1.0<|\eta|<1.5$ and $1.5<|\eta|<2.8$, respectively.

The PHENIX central arms comprise two sets of detectors located on the 
west and east sides of the beam axis. Each arm covers 90$^{\circ}$ in 
azimuth and a pseudorapidity range of $|\eta|<0.35$. Track and momentum 
reconstructions of charged particles were performed with the drift 
chambers (DC) and pad chambers (PC).  The DC are located at a radial 
distance of 2.02 m to 2.46 m from the beam axis in the west and east 
arms, covering 2 m length along the beam axis.  The PC are multi-wire 
proportional chambers in each of the central arms, and are located at 
radial distances of 2.5 m (PC1) and 4.9 m (PC3).  The tracks and momenta 
were reconstructed by combining the hit information in the DC and PC1, 
providing a momentum resolution of $\delta p/p \approx 1.3\% \oplus 1.2\% 
\times p$ GeV/$c$~\cite{momres}.  Global-track reconstruction was 
performed by associating these tracks with hits in the outer detectors, 
such as PC3 and the lead-scintillator (PbSc) electromagnetic calorimeters, 
as shown in Fig.~\ref{fig1}.  Particle identification is provided by 
the PbSc~\cite{EMCal}, which is a sampling calorimeter with a timing 
resolution of about 500 ps~\cite{momres} located at a radial distance 
of 5.1 m from the beam axis.


\section{Data Analysis\label{sec:ana}}

The $\sqsn$ = 200 GeV Au$+$Au collision data were collected by PHENIX 
during the 2007 running period. A total of 4.2 billion events 
were used for this analysis, where the minimum bias trigger 
with at least two hits in each BBC was required. This trigger measures
92$\pm$3\% of the total inelastic cross section~\cite{ppg146}.
Additional offline requirements of one zero-degree-calorimeter hit on each side 
and a collision vertex position of less than $\pm$30 cm were applied.

\subsection{Track Selection}

Charged tracks with good quality were selected based on the track 
information from the DC and PC1. To reduce the background due to the 
random association of hits and reconstructed tracks, track residuals were 
required to be less than 2$\sigma$ in the $\phi-z$ plane at the PC3 and 
PbSc for pions. For kaons, this cut was relaxed to 2.5$\sigma$ to 
increase statistics. The fraction of the random background is $\sim$4.6\% 
after the 2$\sigma$ cuts and $\sim$5.3\% after the 2.5$\sigma$ cuts at 
\pt=0.5 GeV/$c$ in the 0\%--10\% most central collisions. The effect of the 
track quality cuts was included in the systematic uncertainty.

\subsection{Particle Identification}

Particle identification was performed by combining time-of-flight data 
from the PbSc in the west arm, the reconstructed momentum, and flight path 
length from the collision vertex to the hit position at the PbSc wall. The 
squared mass of the particles is given by the following formula: 
\begin{equation} 
m^{2} = \frac{p^{2}}{c^{2}} \left[ \left( \frac{ct}{L} \right)^{2} - 1 \right], 
\end{equation} 
where $p$ is the momentum, $t$ is 
the time-of-flight, $L$ is the flight path length, and $c$ is the speed of 
light. Pions and kaons were selected from a 2$\sigma$ window around their 
peaks in the squared mass distribution. Additional requirements, i.e. to 
be away from the mass peak of other particles, were applied to reduce 
contamination. The $\pi$/$K$ separation was achieved up to a momentum of 
$\sim$1 GeV/$c$. Contamination in the pion samples from kaons is below 1\% 
for $p \approx 1$ GeV/$c$ and contamination in the kaon samples from pions 
(protons) is below 4\% (1\%) for $p \approx 1$ GeV/$c$.

\subsection{Construction of the Correlation Function}

In this section, a bold character denotes four-dimensional vector and an 
arrow denotes three-dimensional vector.

The experimental correlation function defined as
\begin{equation}
C_2({\bf q}) = \frac{A({\bf q})}{B({\bf q})},
\end{equation}
was measured as a function of the pair momentum difference 
${\bf q}={\bf p_1}-{\bf p_2}$, where $A$ and $B$ are constructed from 
identical particle pairs from the same event and mixed-event 
respectively. The mixed-events are taken from similar event 
centralities and vertex positions. In the case of azimuthal-dependent 
analysis, the mixed-events are also required to have similar values for 
the second-order event plane defined in~Sec.\ref{sec:EPdep}.

Particle pairs with similar momenta and spatially close to each other are 
affected by incorrect track reconstruction and detector inefficiencies. 
These effects were removed by applying pair selection cuts at the DC and 
PbSc following our previous analysis~\cite{ppg149,ppg82}. In addition, 
pairs that are associated with hits on the same tower of the PbSc were 
removed.

The particle pairs were analyzed with the Bertsch-Pratt 
parameterization~\cite{ScottPara,BertschPara} as functions of the pair 
momentum difference ${\bf q}$ and mean pair momentum $\vec{k}$, where 
$\vec{k}=(\vec{p_1}+\vec{p_2})/2$. The $\vec{k}$ is projected into its 
longitudinal component $k_{\rm z}$ and transverse component $\vec{k_{\rm 
T}}$. The ${\bf q}$ is projected into the longitudinal (\ql), outward 
(\qo), and sidewards (\qs) components, where \ql denotes the beam 
direction, \qo is perpendicular to $\vec{k_T}$, and \qs is 
perpendicular to both \ql and \qo. In this frame, the energy (temporal) 
component of the four-dimensional vector is taken in the outward component 
by performing the analysis in the longitudinal co-moving system, where 
$k_{\rm z}=0$.

The $C_2({\bf q})$ function is divided into two components based on the 
core-halo picture in which the $\lambda$ parameter controls the relative 
strength of the core and the halo.
\begin{align}
&C_{2}({\bf q}) = C_2^{\ core} + C_2^{\rm halo} \nonumber \\
&\quad\;\;\,\,\hspace{6.5pt} = \lambda (1+G({\bf q})) F_{c}(q) + (1-\lambda) \label{eq:corehalo}\\
&G(\qs,\qo,\ql) = e^{ -R_{\rm s}^{2}q_{\rm s}^{2} -R_{\rm o}^{2}q_{\rm o}^{2} -R_{\rm l}^{2}q_{\rm l}^{2} -2R_{\rm os}^{2}q_{\rm s}q_{\rm o} } \label{eq:BPratt}
\end{align}
The $F_{c}(q)$ is the Coulomb correction factor evaluated by the Coulomb 
wave function~\cite{Bowler, Sinyukov}, where $q$ is the scalar 
quantity of ${\bf q}$. The central core contributes to the quantum statistical 
interference. The halo includes the decay of long-lived particles for 
which the quantum statistical interference occurs in a $q$ range that is 
too small to be resolved experimentally, and for which the Coulomb 
interaction is negligible. The core is assumed to be a Gaussian source as 
given by Eq.~\eqref{eq:BPratt}.

The HBT radii denoted by \Rs, \Ro, and \Rl represent the spatial extent of 
the emission region in each direction, but \Ro and to a lesser extent \Rl 
and \Ros include a contribution from the emission duration. All radii are 
sensitive to position-momentum correlations.  The \Ros term arises in the 
case of azimuthal-dependent analysis due to asymmetries in the emission 
region~\cite{femtoscopy}, while it vanishes in the azimuthal-integrated 
analysis.

The HBT radii were measured as a function of \kt and presented as a 
function of the transverse mass $\mt=(\kt^2+m^2)^{1/2}$ to study 
particle-species dependence, where $m$ is the particle mass.

\subsection{Event Plane Dependence}\label{sec:EPdep}

The second-order event-plane angle ($\Psi_{2}$) was determined using the 
RXNP detector based on the azimuthal anisotropy of emitted particles in 
momentum space:
\begin{eqnarray}
\Psi_2 &=& \frac{1}{2} \tan^{-1} \left( \frac{Q_{y}}{Q_{x}} \right), \\
Q_x &=& \sum w_i \cos(2\phi_i), \\
Q_y &=& \sum w_i \sin(2\phi_i), 
\end{eqnarray}
where $\phi_i$ is the azimuthal angle of each segment $i$ in the RXNP and 
$w_i$ is the weight which reflects the particle multiplicity in that 
segment. Corrections for detector acceptance as detailed in 
Ref.~\cite{v2sys_phnx} were applied.

Due to the finite number of particles within the RXNP acceptance, the 
observed event plane $\Psi_2$ is smeared around the true event plane 
$\Phi$. This smearing effect is typically accounted for by the resolution. 
The event plane resolution defined as ${\rm Res}\{\Psi_2\}=\langle 
\cos[2(\Psi_{2}-\Phi)] \rangle$ was estimated by the two-subevent 
method~\cite{TwoSub} using the event plane correlation between the RXNP at 
forward and backward angles. The ${\rm Res}\{\Psi_{2}\}$ has a maximum of 
0.75 in midcentral events~\cite{momres}.

The finite event plane resolution reduces the oscillation amplitude of HBT 
radii relative to the event plane. To take this effect into account, a 
model-independent correction suggested in Ref.~\cite{bw_hbt} was applied. 
The pair distribution measured at a certain azimuthal angle $\phi$ 
relative to the reconstructed event plane, $N({\bf q},\phi-\Psi_2)$, is 
smeared by the finite event plane resolution and finite width of angular 
bins $\Delta$. The Fourier coefficients for the true and measured $N({\bf 
q},\phi-\Psi_2)$, $N_{\alpha,n}({\bf q},\phi-\Psi_2)$, can be associated 
with the following relation:
\begin{eqnarray}
N^{\rm exp}_{\alpha,n}({\bf q},\phi-\Psi_2) &=& N^{\rm 
true}_{\alpha,n}({\bf q},\phi-\Phi) \frac{\sin(n\Delta/2)}{n\Delta/2} \\ \nonumber
&\times& \langle \cos[n(\Psi_{2}-\Phi)] \rangle,
\label{eq:EPcr_fc}
\end{eqnarray}
where $\alpha$ denotes sine and cosine terms of the Fourier coefficients 
($\alpha=s, c$) and $n$ denotes the order of the coefficient. The above equation is analogous to the correction for the 
elliptic flow ($v_2$). Based on Eq.~\eqref{eq:EPcr_fc}, the $A({\bf q})$ and 
$B({\bf q})$ functions can be unfolded by using the following equation:
\begin{eqnarray}
N(q,\phi _{j}) &=& N_{\rm exp}(q,\phi_j) + 2 \sum _{n=m, 2m, \cdots}^{n_{\rm bin}/2} \zeta _{n,m}(\Delta ) \nonumber\\
&\times& [N_{c,n}^{\rm exp}(q) \cos(n \phi_j) + N_{s,n}^{\rm exp}(q) \sin(n \phi_j)],\nonumber\\
\label{eq:rp_crr}
\end{eqnarray}
where $n_{\rm bin}$ is the number of azimuthal angular bins, and $m$ is the 
order of the event plane, and $\phi_j$ denotes the center of $j^{th}$ 
angular bin which corresponds to the azimuthal angle of the pair with 
respect to the event plane. $N_{c,n}^{\rm exp}(q)$, $N_{s,n}^{\rm exp}(q)$, and 
$\zeta_{n,m}(\Delta)$ are given by
\begin{eqnarray}
N_{c,n}^{\rm exp}(q) &=& \langle N_{\rm exp}(q,\phi-\Psi_2 ) \cos [n (\phi-\Psi_2) ] \rangle, \nonumber \\
                 &=& \sum_j N_{\rm exp}(q,\phi_j ) \cos (n \phi_j ) / n_{\rm bin}, \\
N_{s,n}^{\rm exp}(q) &=& \langle N_{\rm exp}(q,\phi-\Psi_2 ) \sin [n (\phi-\Psi_2) ] \rangle, \nonumber \\
                 &=& \sum_j N_{\rm exp}(q,\phi_j ) \sin (n \phi_j ) / n_{\rm bin}, \\
\zeta _{n,m}(\Delta ) &=& \frac{n \Delta /2}{\sin (n \Delta /2) \langle \cos[n(\Psi_{m}^{\rm obs}-\Phi)] \rangle} - 1. \label{eq:zeta}
\end{eqnarray}
The details of Eqs.~\eqref{eq:EPcr_fc}--\eqref{eq:zeta} can be found in
Ref.~\cite{bw_hbt}.

\subsection{Systematic Uncertainties}\label{sec:SE}

Systematic uncertainties were estimated by variations of track quality 
cuts at PC3 and PbSc, pair selection cuts, and particle identification 
(PID) cuts. Also, the effect of the Coulomb correction was studied by 
varying the input source size in the calculation of $F_{c}(q)$ in 
Eq.~\eqref{eq:corehalo}. Typical systematic uncertainties of the measured 
radii for charged pions and kaons are listed in Table~\ref{table_SE_pi} 
and Table~\ref{table_SE_k}.

In the azimuthal-dependent analysis, the variations when using different 
event planes from forward, backward, and both combined RXNPs, were also 
incorporated. The systematic uncertainties of the oscillation amplitudes 
of HBT radii were dominated by the event plane determination, which were 
16\% on average in the final eccentricity defined by the oscillation of 
$\Rs^{2}$ and the same fraction of the uncertainty was assumed for pions 
and kaons.

\begin{table}[ht]
\caption{Typical systematic uncertainties of HBT parameters for positive 
pion pairs in 0\%--10\% centrality and $0.6<\kt<0.7$ GeV/$c$.}
\begin{ruledtabular}\begin{tabular}{ccccc}
systematic source & $\lambda$ & $R_{\rm s}$ & $R_{\rm o}$ & $R_{\rm l}$ \\ 
                  &      & [\%] & [\%] & [\%] \\ 
\hline
track quality   & 1.8  & 0.3  & 0.5  & 3.1 \\
pair selection  & 4.3  & 1.0  & 4.6  & 3.7 \\
Particle ID     & 0.4  & 0.3  & 1.3  & 0.0 \\
Coulomb         & 0.4  & 0.1  & 0.3  & 0.1 \\ 
\\
Total           & 4.7  & 1.1  & 4.8  & 4.8 \\ 
 \end{tabular}\end{ruledtabular}
\label{table_SE_pi}
\end{table}

\begin{table}[ht]
\caption{Typical systematic uncertainties of HBT parameters for charge 
combined kaon pairs in 0\%--10\% centrality and $0.3<\kt<0.68$ GeV/$c$.}
\begin{ruledtabular}\begin{tabular}{ccccc}
systematic source      & $\lambda$ & $R_{\rm s}$ & $R_{\rm o}$ & $R_{\rm l}$ \\ 
                       &      & [\%] & [\%] & [\%] \\ 
\hline
track quality    & 5.1 & 2.2 & 1.9 & 2.2 \\
pair selection   & 9.0 & 1.5 & 0.1 & 1.9 \\
Particle ID              & 6.1 & 0.3 & 4.5 & 0.1 \\
Coulomb          & 4.6 & 0.3 & 1.1 & 0.2 \\ 
\\
Total            & 12.9& 2.7 & 5.0 & 2.9 \\ 
\end{tabular}\end{ruledtabular}
\label{table_SE_k}
\end{table}

The effect of momentum resolution was studied employing the same method as 
previous analyses~\cite{BE_NA49,star_pionhbt}. The momentum was smeared 
according to the known momentum resolution and the correlation function 
was reconstructed using the smeared $A({\bf q}$) and $B({\bf q}$) 
functions. By taking the ratio of the smeared and unsmeared correlation 
function, the correction factor was obtained. The correction on the 
momentum resolution was performed by multiplying the correction factor to 
the measured correlation function. The correction did not affect \Rs and 
\Rl, but slightly increased $\lambda$ ($<$10\%) and \Ro ($<$6\%).


\section{Results and Discussion\label{sec:result}}

\subsection{Azimuthal-integrated analysis}

\begin{figure}[htb]
\includegraphics[width=0.998\linewidth]{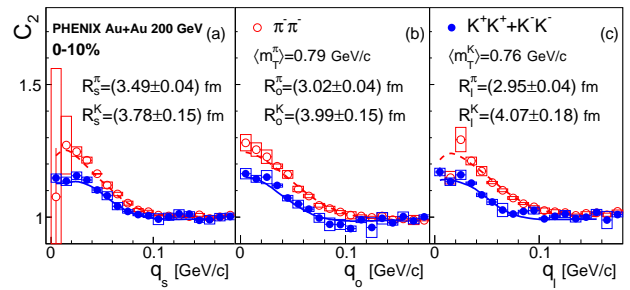}
\caption{(Color online) 
Correlation functions of negative pions and charged kaons for 0\%--10\% 
centrality (a-c), where positive and negative kaons are combined. Open boxes show the systematic uncertainties.
Solid and dashed lines show the fit functions and the extracted radii values are 
shown in the figure. }
\label{fig2}
\end{figure}

Figure~\ref{fig2}(a-c) shows an example of correlation functions of pion 
pairs and kaon pairs in 0\%--10\% centrality in a \kt bin with fit lines 
given by Eq.~\eqref{eq:corehalo}, where the momentum correction is not 
applied. The \kt range is selected to have similar \mt for pions and 
kaons. The 3-dimensional $A({\bf q})$ and $B({\bf q})$ functions are 
projected in each 
${\bf q}$ direction. In the projection, the other $q$ are restricted to be 
less
than 40 MeV/$c$ (e.g. when making $C_{2}(\qs)$, the projection ranges of 
\qo and \ql should be \qo$<$40 MeV and \ql$<$40 MeV). The 1D correlation 
functions shown in Fig.~\ref{fig2} are obtained by taking ratio of the 
projected $A({\bf q})$ and $B({\bf q})$ functions. The extracted HBT radii 
with the statistical uncertainties are also shown in each panel. The width of the 
enhancement at the low ${\bf q}$ region in the correlation function is 
proportional to the inverse of the HBT radius. The width of the 
correlation function is comparable between pions and kaons in the sidewards 
direction, but narrower for kaons in the outward and longitudinal 
directions, indicating a larger radius in those directions than for pions 
with a similar mean \mt. We note that the data points at lower q bins
fluctuate due to lower statistics. The effect of the fluctuation on the radii was studied by
varying the fit range and it was found to be within a few percent for both pions and kaons.

\begin{figure*}[htb]
\includegraphics[width=0.998\linewidth]{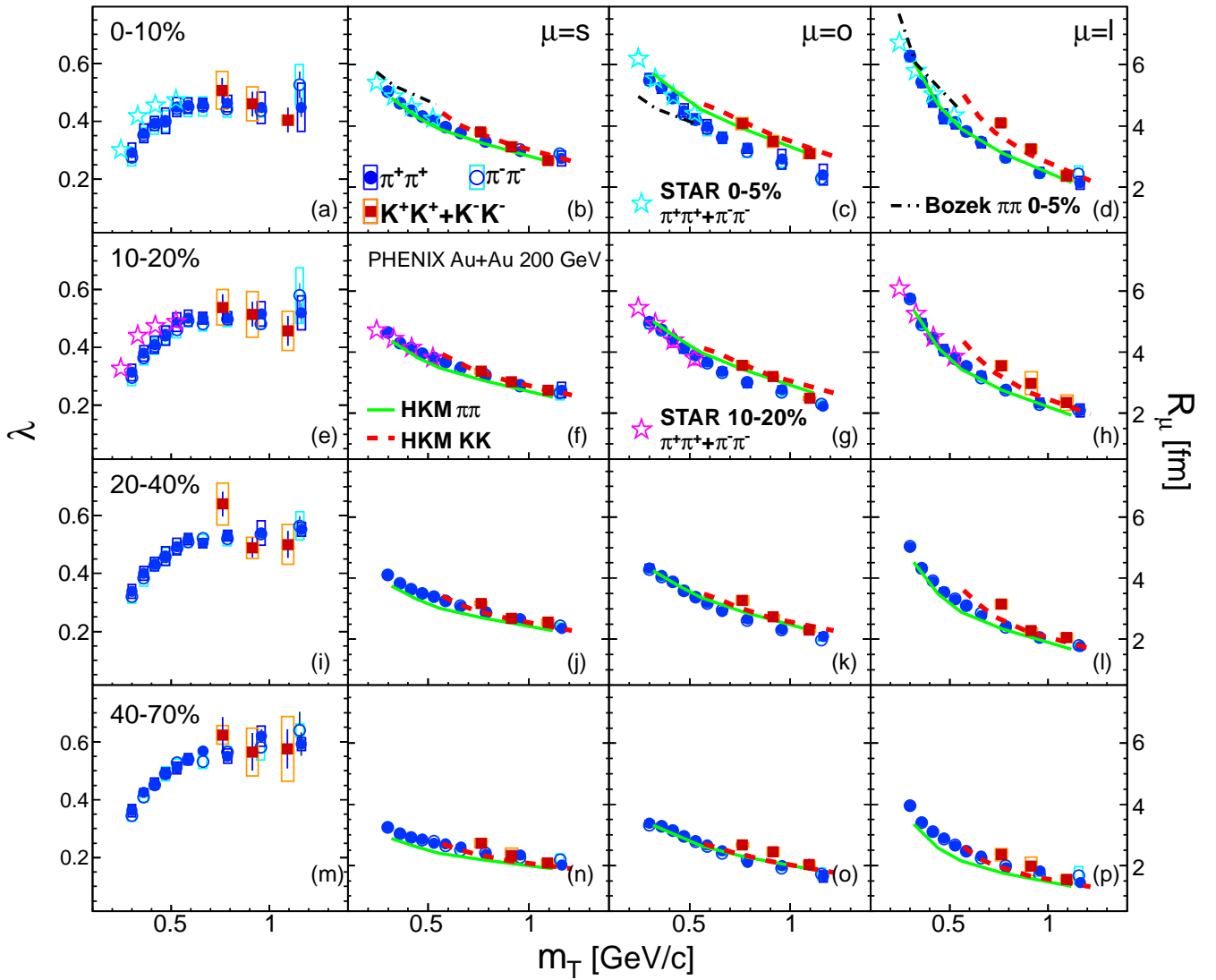}
\caption{(Color online) 
Extracted HBT parameters of charged pions and kaons as a function of \mt 
for the centralities indicated, where open boxes show the systematic 
uncertainties. Results of charged pions from STAR~\cite{star_pionhbt} are 
compared. Calculations from the hydrokinetic model 
(HKM)~\cite{HKM_compared} and viscous-hydrodynamic model 
(Bozek)~\cite{Bozek} are also shown.}
\label{fig3}
\end{figure*}

\begin{figure*}[htb]
\includegraphics[width=0.998\linewidth]{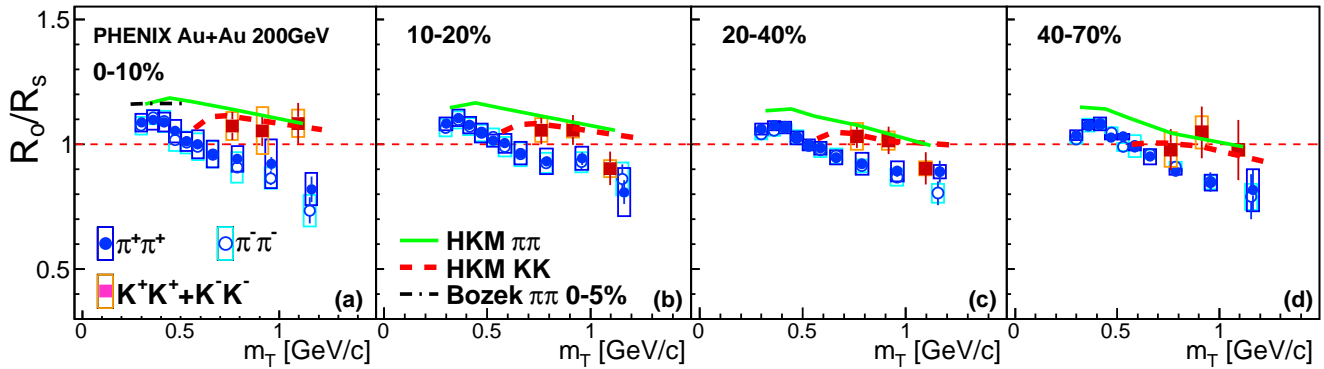}
\caption{(Color online) 
Ratio of \Ro and \Rs for charged pions and kaons as a function of \mt. 
Calculations from the hydrokinetic model (HKM)~\cite{HKM_compared} and 
viscous-hydrodynamic model (Bozek)~\cite{Bozek} are also shown.}
\label{fig4}
\end{figure*}

\begin{figure*}[htb]
\includegraphics[width=0.998\linewidth]{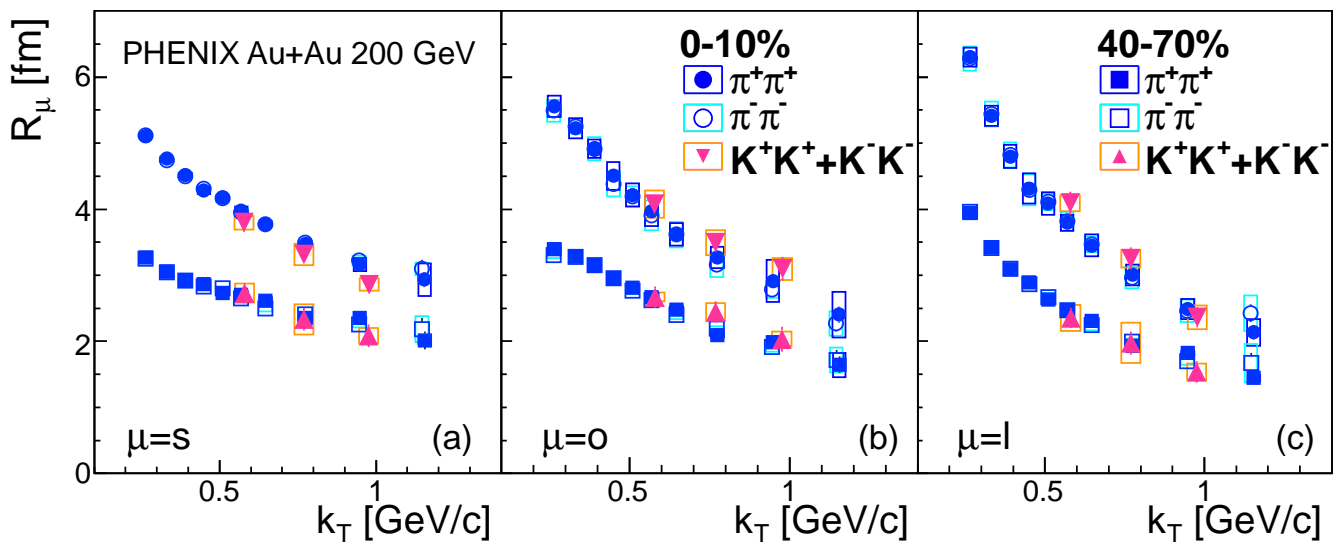}
\caption{(Color online) 
HBT radii of charged pions and kaons as a function of \kt, where open 
boxes show the systematic uncertainties.}
\label{fig5}
\end{figure*}

\begin{figure}[htb]
\includegraphics[width=1.0\linewidth]{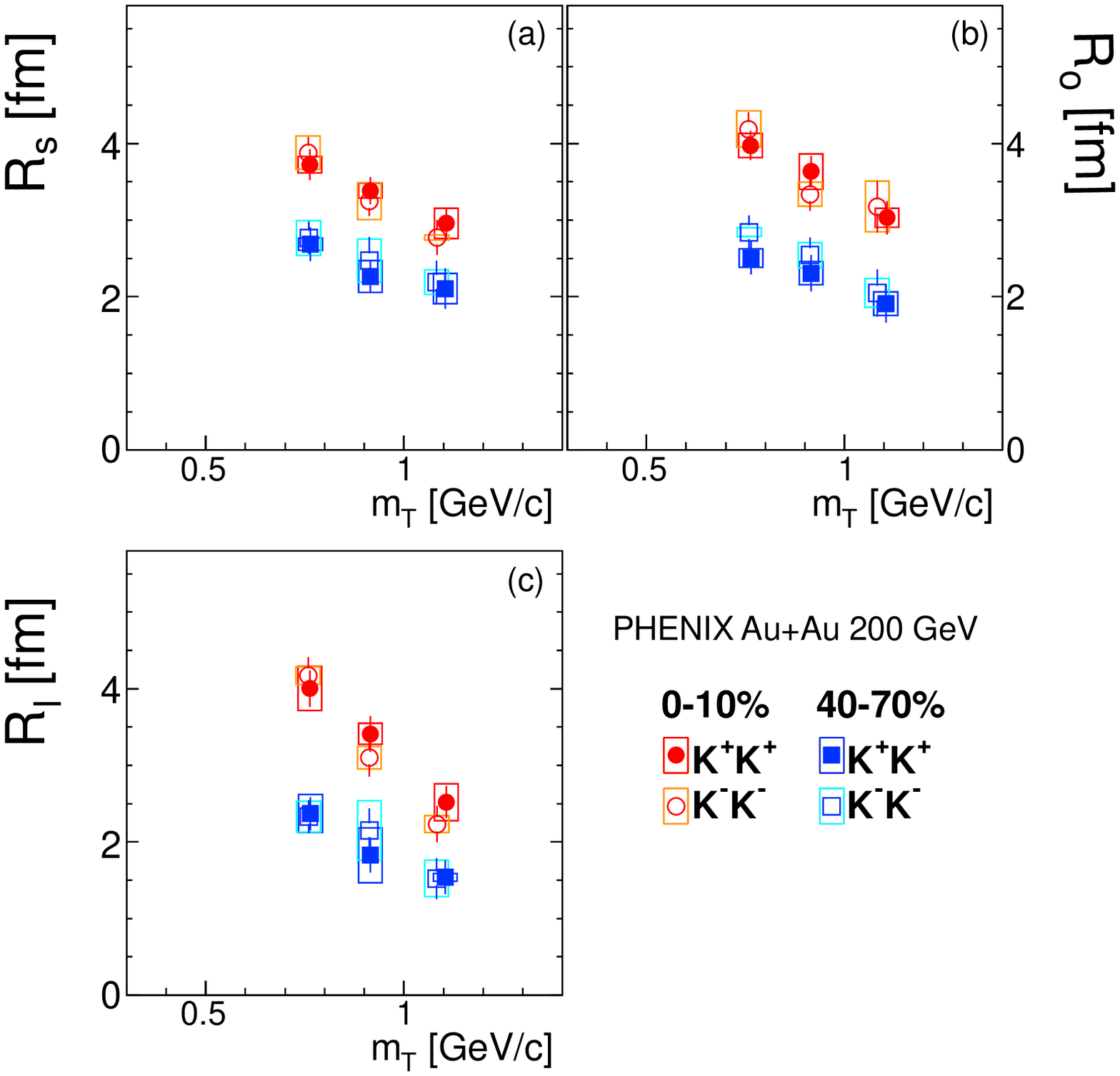}
\caption{(Color online) 
Comparison of HBT radii between positive and negative kaons pairs in 
central and peripheral collisions, where open boxes show the systematic 
uncertainties.}
\label{fig6}

\includegraphics[width=0.998\linewidth]{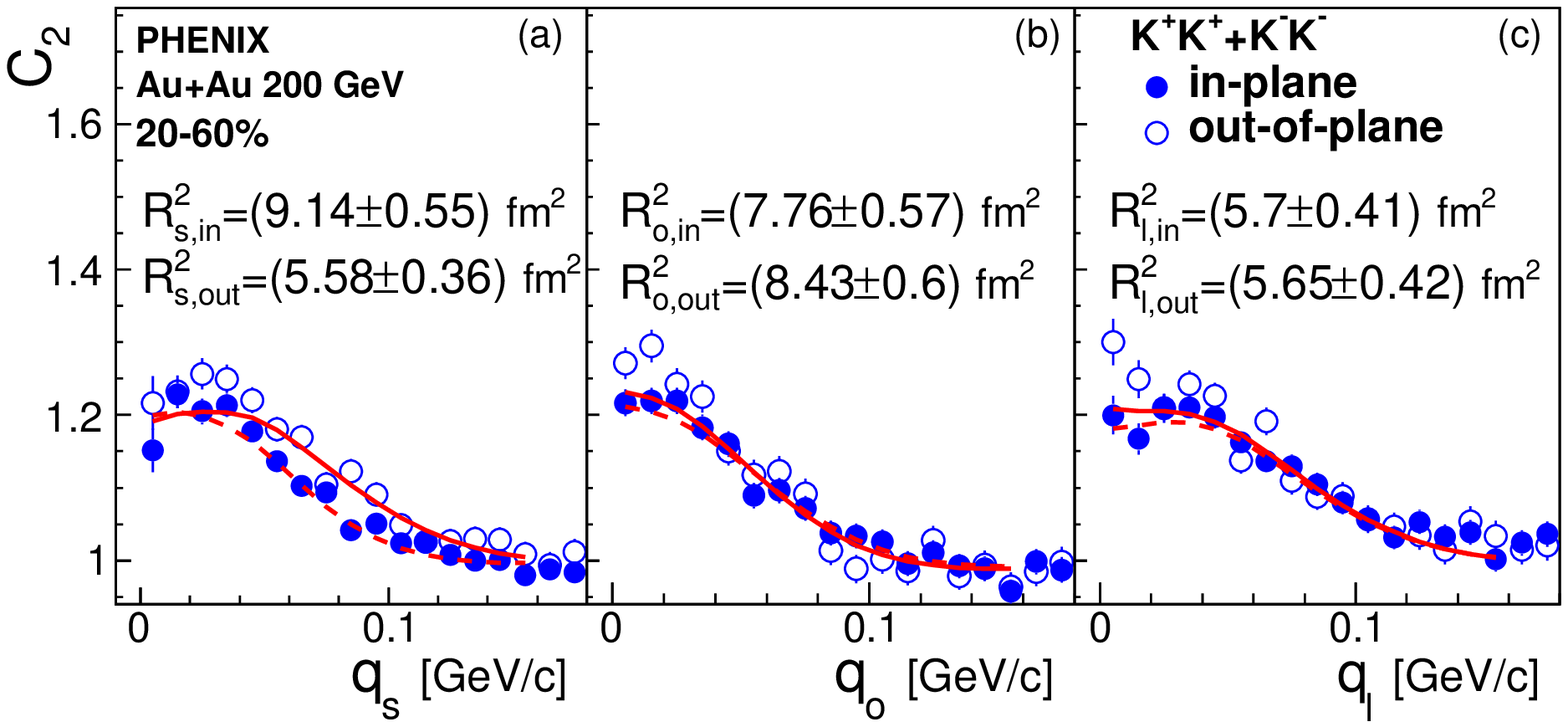}
\caption{(Color online) Correlation functions of charged kaons in 
20\%--60\% centrality (a-c), where positive and negative kaons are 
combined. The correlation functions along \qs and \qo directions are 
averaged out between positive and negative $q$. Lines show the fit 
functions and the extracted radii values are shown in the figure.}
\label{fig7}
\end{figure}

\begin{figure}[htb]
\includegraphics[width=1.0\linewidth]{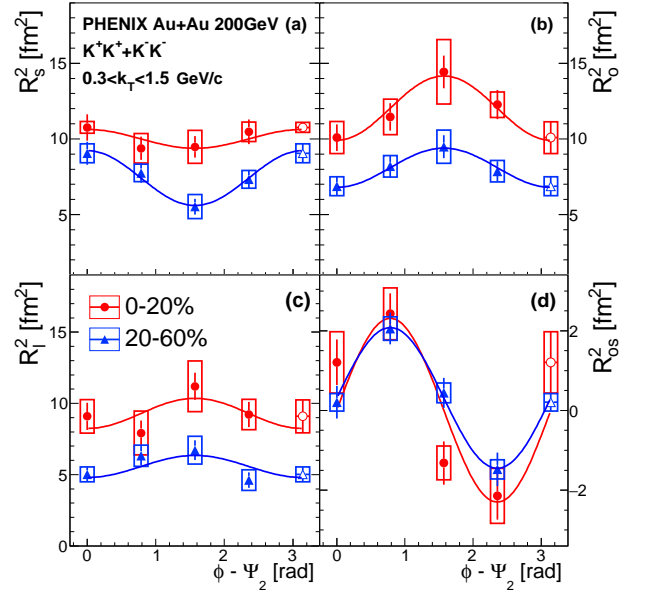}
\caption{(Color online) 
Squared HBT radii of charged kaon pairs as a function of azimuthal pair 
angle $\phi$ with respect to $\Psi_2$ for two centrality bins, where \kt 
is integrated over 0.3-1.5 GeV/$c$. The open symbols at $\phi-\Psi_2=\pi$ 
are the same data as that at $\phi-\Psi_2=0$. Open boxes show systematic 
uncertainties and the solid lines are the fitting functions given by 
Eq.~\eqref{eq:cosine_series}.}
\label{fig8}
\end{figure}

\begin{figure*}[bth]
\includegraphics[width=0.998\linewidth]{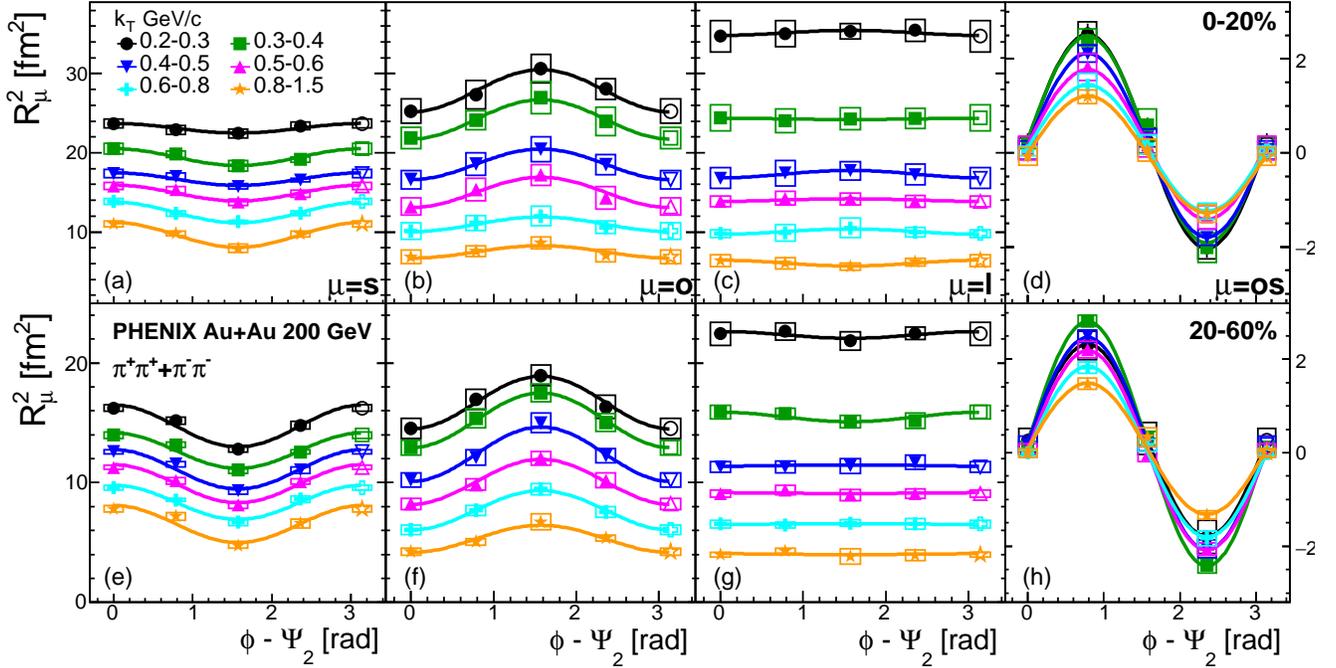}
\caption{(Color online) 
Squared HBT radii of charged pion pairs as a function of azimuthal pair 
angle $\phi$ with respect to $\Psi_2$ for 6 \kt bins and 2 centrality bins 
((a)-(d) for 0\%--20\% and (e)-(h) for 20\%--60\%), where open symbols at 
$\phi-\Psi_2=\pi$ is the same data as that at $\phi-\Psi_2=0$. Open boxes 
show systematic uncertainties and solid lines show the fit functions by 
Eq.~\eqref{eq:cosine_series}.}
\label{fig9}
\end{figure*}

\begin{figure*}[htb]
\includegraphics[width=0.998\linewidth]{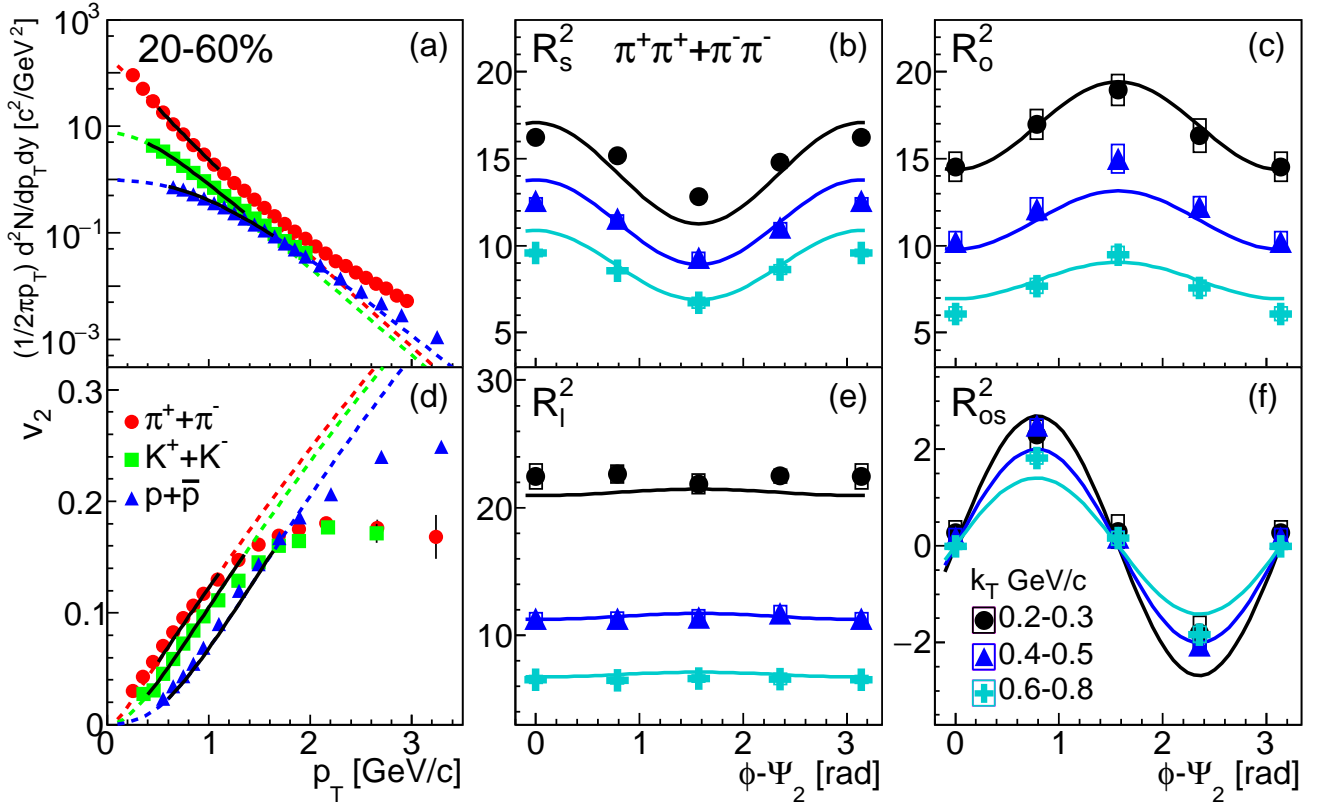}
\caption{(Color online) 
The blast-wave model fits (Fit A) to the \pt spectra (a), elliptic flow of 
$\pi$, K, p (d), and HBT radii of $\pi$ (b,c,e,f). The solid lines show 
the fit functions.}
\label{fig10}
\end{figure*}

\begin{figure}[htb]
\includegraphics[width=1.0\linewidth]{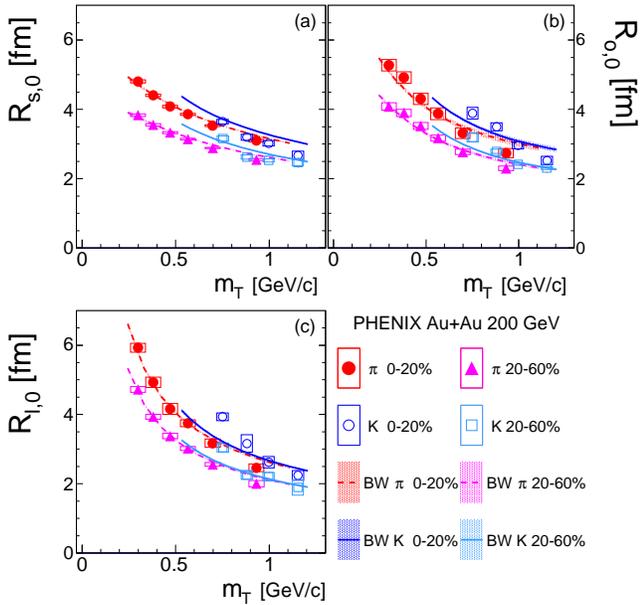}
\caption{(Color online) 
The 0$^{\rm th}$-order Fourier coefficient of HBT radii of charged pions 
as a function of \mt, and blast-wave model calculations for both pions 
(dashed line) and kaons (solid line), where the blast-wave model 
parameters shown in Fit A of Table~\ref{table_bwpara} were used.}
\label{fig11}
\end{figure}

\begin{figure*}[tbh]
\includegraphics[width=0.998\linewidth]{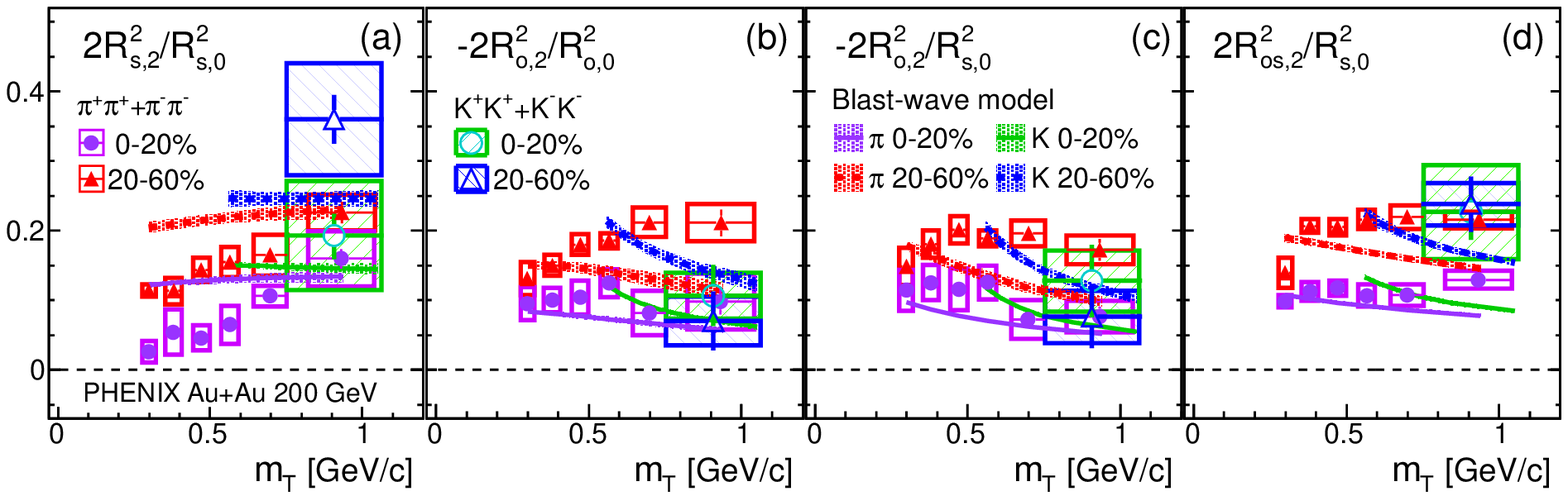}
\caption{(Color online) 
Oscillation amplitudes relative to the average for four different 
combinations of the azimuthal-dependent HBT radii as a function of \mt for 
charged pions and kaons. Open boxes show systematic uncertainties. 
Calculations from the blast-wave model with parameters of Fit A shown in 
Table~\ref{table_bwpara} are shown for comparison.}
\label{fig12}

\includegraphics[width=0.998\linewidth]{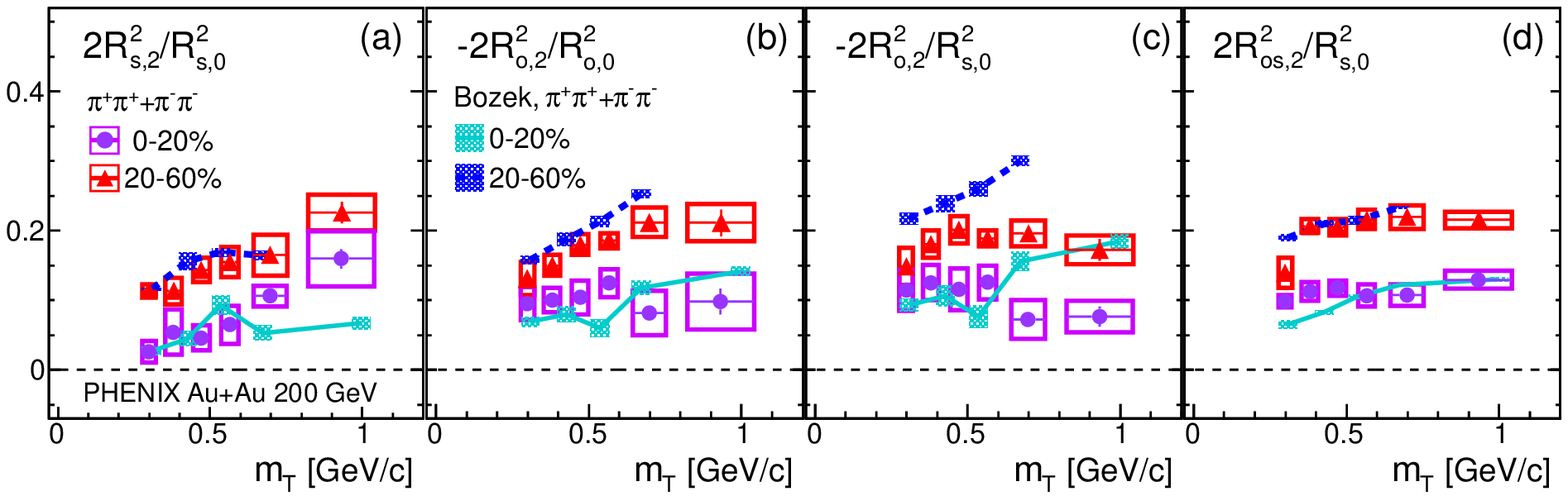}
\caption{(Color online) 
Oscillation amplitudes relative to the average for four different 
combinations of the azimuthal-dependent HBT radii as a function of \mt for 
charged pions. Open boxes show systematic uncertainties. Calculations from 
the 3+1-D viscous-hydrodynamic model~\cite{Bozek} are shown for 
comparison.}
\label{fig13}
\end{figure*}

Figure~\ref{fig3} shows the extracted HBT parameters of charged pions and 
kaons for four centrality classes as a function of \mt. Results for 
charged pions in the low \mt region from STAR~\cite{star_pionhbt} 
are also plotted. The source parameters 
from the two experiments are in good agreement, but the $\lambda$ 
parameters are 20\% lower at low \mt. The value of $\lambda$ is sensitive 
to the combinatorial background level, which may differ between PHENIX and 
STAR. Positive and negative pions are quite consistent. The presented data 
are also consistent with our previous results~\cite{pihbt,ppg82}.

The decrease of HBT radii with \mt is often attributed to the 
position-momentum correlation induced by collective flow. The slope of the 
\mt dependence becomes steeper for more central collisions, which is 
consistent with an expectation of a stronger radial 
flow~\cite{femtoscopy}. \Rs shows approximate \mt scaling between pions 
and kaons as predicted by the Buda-Lund model~\cite{Buda}, which is based 
on the analytic approach of the perfect fluid hydrodynamics. On the other 
hand, \Ro and \Rl of kaons show larger values than those of pions as noted 
already in Fig.~\ref{fig2}, where the \mt scaling is broken. The 
difference increases with centrality going from peripheral to central 
collisions. The similar difference between pions and kaons for \Rl was 
reported by STAR~\cite{k3Dsource}.

The results are compared with the hydrokinetic model 
(HKM)~\cite{HKM,HKM_compared}. The HKM incorporates realistic conditions 
such as the Glauber initial condition, crossover transition, fluid 
hydrodynamics, microscopic transport and resonance decays, but does not 
explicitly include the viscous correction. It is reported that the model 
calculations with the initial condition of the color glass condensate are 
very similar to those with the Glauber initial condition~\cite{HKM}. As 
shown in Fig.~\ref{fig3}, the HKM~\cite{HKM_compared} describes well the 
overall trend of HBT radii for pions and kaons in all centrality bins, 
however it overestimates \Ro of pions in more central collisions and 
underestimates \Rs and \Rl of pions in peripheral collisions. The HKM also 
describes the difference of pions and kaons in the longitudinal direction, 
which can be understood by strong transverse flow~\cite{HKM_lhc}, but the 
difference in the outward direction cannot be explained well. The data for 
pions in most central collisions are also compared with (3+1)-D viscous 
hydrodynamic model~\cite{Bozek} calculations which employ a Glauber 
initial condition and $\eta/s=0.08$ (also see Sec.~\ref{sec:E2vsHydro} for 
details). The model follows the general trends in the data.

The ratio of \Ro and \Rs, which is sensitive to the emission duration of 
particles, is also plotted as a function of \mt in Fig.~\ref{fig4}.  
Results for both species do not show any significant centrality 
dependence, but the values for kaons are larger than those for pions at 
all \mt and centralities, a possible indication of longer emission 
duration time for kaons than for pions. The HKM reproduces the data for 
kaons well, but not for pions.

The \mt scaling of HBT radii was inspired by the hydrodynamic 
expansion~\cite{mTscaling}. This is based on the idea that the kinetic 
freeze out of hadrons occurs at the same time and the hadrons with similar 
velocities are emitted from the same homogeneity region. In other words, 
the homogeneity length depends on the particle mass under the presence of 
radial flow. In Fig.~\ref{fig5}, both pion and kaon HBT radii for central 
and peripheral events are plotted as a function of \kt. Unlike the case of 
the \mt dependence shown in Fig.\ref{fig3}, both radii seem to be scaled 
better for \kt in all $q$ directions as predicted in Ref.~\cite{HKM_lhc}. 
This model includes many different effects such as the hadronic cascade 
and resonance decays in addition to radial flow.

We have also checked charge-dependent kaon HBT radii in Fig.~\ref{fig6}. 
There was no significant difference between positive and negative kaons as 
we expected. If nucleons are dominant in the particle-emitting source and 
the net baryon density is not small, the measured radii might be different 
between $K^{+}$ and $K^{-}$ (and also pions) because of smaller cross 
section of $K^{+}-N$ than $K^{-}-N$~\cite{Beringer:1900zz}. However this is not 
observed.

\subsection{Azimuthal-dependent analysis}

\subsubsection{Results}

We have measured the azimuthal angle dependence of HBT radii with respect 
to $\Psi_2$ for both charged pions and kaons. Figure~\ref{fig7}(a-c) shows 
the correlation functions of charged kaons in the 20\%--60\% centrality bin 
in the in-plane ($|\phi-\Psi_2|<\pi/16$) and out-of-plane 
($|\phi-\Psi_2-\pi/2|<\pi/16$) directions without correction for the event 
plane resolution. The correlation functions in Fig.~\ref{fig7} are 
calculated in the same way as Fig.~\ref{fig2}, i.e. when making the 1-dimensional $C_{2}$ along the q of interest, 
the other $q$ are limited to be less than 50 MeV/$c$.
To make a comparison of the $C_{2}$ width between in-plane and out-of-plane directions,
the $C_{2}$ in the positive and negative $q$ are averaged because they are symmetric over $q=0$ within the statistical uncertainties.
The extracted radii without the correction are also shown in the figure. A difference of the width in 
the correlation function between these (in- and out-of-plane) directions 
can be seen in the sidewards (Fig.~\ref{fig7}(a)) direction. 
Figure~\ref{fig8} shows the extracted HBT radii of charged kaons as a 
function of azimuthal pair angle $\phi$ with respect to $\Psi_2$ for two 
centrality bins where $\langle \kt \rangle$ is $\sim$0.77 GeV/$c$. We 
first fix $\lambda$ in Eq.~\eqref{eq:corehalo} by taking the average for 
$\lambda$ obtained in all azimuthal bins, then we fit in individual 
azimuthal bins again with fixed $\lambda$ parameter as detailed 
in~\cite{star_pionhbt}. This treatment is based on the assumption that 
$\lambda$ has no azimuthal angle dependence, and the data fluctuate but
do not depend on the azimuthal angle beyond the systematic uncertainty.
The cosine oscillations of 
$\Rs^{2}$ and $\Ro^{2}$ (Fig.~\ref{fig8}(a,b)) and the sine oscillation of 
$\Ros^{2}$ (Fig.~\ref{fig8}(d)) can be clearly seen. Non-zero $\Ros$ at 
$(\phi-\Psi_2)=\frac{1}{4}\pi, \frac{3}{4}\pi$ implies that the direction 
of the particle 
emission is tilted relative to the main axis of the emission region. The 
oscillation of $\Rs^{2}$ seems to be larger than for $\Ro^{2}$ in 20\%--60\% 
centrality bin.

We have also measured the charged pion HBT radii with respect to $\Psi_2$ 
for the same centrality bins as kaons with six \kt (\mt) bins as shown in 
Fig.~\ref{fig9}. The averages of $\Rs^{2}$, $\Ro^{2}$, and $\Rl^{2}$ 
decrease with \kt as seen in Fig.~\ref{fig3}. The $\Rs^{2}$ and $\Ro^{2}$ 
have similar but opposite oscillations in all \kt bins. For 20\%--60\% 
centrality, both transverse radii show finite oscillation even in the 
lowest \kt bin, which indicates that the pion emission happens from an 
elliptical source. For 0\%--20\% centrality, the $\Rs^{2}$ has a weak 
azimuthal angle dependence, while the $\Ro^{2}$ has a larger oscillation 
than the $\Rs^{2}$. It could be consistent with $\Ro^{2}$ being more 
influenced by the anisotropic flow as discussed in our previous 
publication~\cite{ppg149}. The oscillations of $\Ros^{2}$ decrease with 
\kt and $\Rl$ displays a negligible azimuthal angle dependence, which 
qualitatively agree with hydrodynamic 
calculations~\cite{bw_hbt,hydro_asHBT2}.

The data shown in Fig.~\ref{fig8} and Fig.~\ref{fig9} are fitted with the 
functions below to extract the oscillation strength~\cite{EpResCrr_hbt}:
\begin{eqnarray}
R_{\rm \mu}^{2} (\Delta\phi) \!\!&=&\!\! R_{\rm \mu,0}^{2} + 2R_{\rm \mu,2}^{2} \cos( 2 \Delta\phi ) \,\,\; (\mu=s, o, l), \nonumber \\
R_{\rm \mu}^{2} (\Delta\phi) \!\!&=&\!\! 2R_{\rm \mu,2}^{2} \sin( 2 \Delta\phi ) \,\,\; (\mu=os), 
\label{eq:cosine_series}
\end{eqnarray}
where $R_{\rm \mu,2}^{2}$ are the 2$^{\rm nd}$-order Fourier coefficient 
and $\Delta\phi=\phi-\Psi_2$. Detailed discussion on the oscillation 
amplitudes is presented in Sec.~\ref{sec:E2vsHydro}.

\subsubsection{Blast-wave model fit}

In this section, we perform blast-wave model fits to our results to 
extract features at the kinetic freeze out and study their particle 
species dependence. The blast-wave (BW) model~\cite{BW_org} is based on a 
hydrodynamical model parameterized by the freeze-out conditions, such as 
the freeze-out temperature ($T_f$) and the transverse flow rapidity 
($\rho_0$). This model is further expanded in Ref.~\cite{bw_hbt} to 
describe the elliptic flow and azimuthal angle dependence of HBT radii by 
introducing additional parameters: 2$^{\rm nd}$-order modulation in 
transverse flow rapidity ($\rho_2$), the transverse source size ($R_{\rm 
x}$, $R_{\rm y}$), the freeze-out time ($\tau_0$), and the emission 
duration ($\Delta\tau$). Once the above seven parameters are fixed, the 
\pt spectra, elliptic flow, and HBT radii can be calculated within the 
model.

Each freeze-out parameter has a different sensitivity to each experimental 
observable~\cite{bw_hbt}. For example, the $\rho_{2}$ and the ratio of 
$R_{\rm x}$ and $R_{\rm y}$ are less sensitive to \pt spectra, but more 
sensitive to the elliptic flow and the azimuthal angle dependence of HBT 
radii. To effectively constrain those parameters, a fit to \pt 
spectra was first performed to determine $T_{f}$ and $\rho_{0}$, then the 
other parameters were determined by simultaneous fit to the elliptic flow 
and azimuthal-dependent HBT radii. For the source size parameters, the 
$R_{\rm x}$ and $R_{\rm y}/R_{\rm x}$ are actually used as the fitting 
parameters.

The BW model assumes that the freeze out for all hadron species takes 
place at the same time, but the actual situation may be more complicated. 
To investigate how the extracted freeze-out parameters vary by 
particle species of HBT radii, the following fits were tested:
\begin{itemize}
\item[A.] Fit for \pt spectra and $v_2$ of $\pi$, $K$, $p$ along with HBT radii of $\pi$.
\item[B.] Fit for \pt spectra and $v_2$ of $\pi$, $K$, $p$ along with HBT radii of $K$.
\end{itemize}
In the case of Fit B, both azimuthal-dependent and azimuthal-integrated 
HBT radii of charged kaons were included in the fit.

Figure~\ref{fig10} shows the results of Fit A for the \pt spectra 
(Fig.~\ref{fig10}(a)) and elliptic flow (Fig.~\ref{fig10}(d)) of $\pi$, 
$K$ and $p$, and pion HBT radii (Fig.~\ref{fig10}(b,c,e,f)) in 20\%--60\% 
centrality. The solid lines show the BW fitting functions and the dashed 
lines in the panel (a) and (d) represent the extended fitting function 
beyond their actual fit ranges. Only three \kt bins for HBT radii are 
shown here, but all six \kt bins shown in Fig.~\ref{fig9} were 
simultaneously used in the fit. Here, the data of \pt spectra are taken 
from~\cite{Spectra_phenix} and the data of $v_2$ from~\cite{ppg124}. The 
\pt spectra and $v_2$ are well described at low \pt, and the overall trend 
of the HBT radii is also reproduced by the BW model.

The results from the fits are summarized in Table~\ref{table_bwpara}. The 
systematic uncertainties of the BW model fit were estimated by varying the 
fit conditions: the fit range, a surface diffuseness to control the 
density profile~\cite{bw_hbt}, and the relative weighting factor between 
different particle species. The systematic uncertainties of data were 
taken into account in the calculation of $\chi^{2}$. Our results from Fit 
A are in good agreement with those in previous 
studies~\cite{Spectra_bw,star_pionhbt}. The results of Fit B shows 
slightly different values, i.e, smaller $R_{\rm x}$ and larger 
$\Delta\tau$. The smaller source ($R_{\rm x}$) might be intuitively 
understood as due to kaons freezing out earlier than pions, but it is not 
significant in the parameter $\tau$. The $\Delta \tau$ obtained by the Fit 
B using the kaon HBT result shows relatively larger values than the 
results by Fit A, which is consistent with the result from $\Ro/\Rs$ as 
shown in Fig.~\ref{fig4}.

Also, the \mt dependence of the pion and kaon HBT radii has been 
calculated using the parameters obtained from Fit A as shown with lines in 
Fig.~\ref{fig11}. For a comparison, the 0$^{\rm th}$-order Fourier 
coefficients ($R_{\rm \mu,0}$) for pions which correspond to the HBT radii 
obtained in the azimuthal-integrated analysis are plotted as filled symbols. 
The kaon HBT radii from the azimuthal-integrated analysis are also compared in the figure.
The BW model shows the $\pi$/$K$ difference in the sidewards and outward 
directions, but not in the longitudinal direction unlike the experimental data.

\begin{table*}[th]
\caption{Summary of extracted parameters in the blast-wave model fit for 
two fitting conditions (see the text for details). The $T_{f}$ and 
$\rho_{0}$ parameters were obtained by fits to \pt spectra, and the other 
parameters were obtained by a simultaneous fit to $v_{2}$ and the HBT 
radii. The values in parentheses represent the systematic uncertainties 
derived by varying the model fit conditions.}
\begin{ruledtabular}\begin{tabular}{cccccccccccc}
Fit & Centrality & $T_{f}$ & $\rho_{0}$ & $\rho_{2}$ & $R_{\rm x}$ & $R_{\rm y}/R_{\rm x}$ & $\tau$ & $\Delta\tau$ & $\chi^{2}$/NDF & $\chi^{2}$/NDF & $\chi^{2}$/NDF \\ 
&    & [MeV] & & & [fm] & & [fm/$c$] & [fm/$c$] & (spectra) & ($v_{2}$) & (HBT) \\ \hline 
A &  0\%--20\%	&	104	&	0.995	&	0.047	&	11.28	&	1.092	&	8.22	&	2.06	&	143.4/27=5.3	&	25.2/21=1.2 	& 2526.7/96=26.3 \\
& &	(5)	&	(0.055)	&	(0.005)	&	(0.23)	&	(0.003)	&	(0.23)	&	(0.18)	&	&	\\
A & 20\%--60\%	&	113	&	0.905	&	0.074	&	8.25	&	1.171	&	6.1	&	1.56	&	206.5/27=7.6	&	46.4/21=2.2 	& 	1998.3/96=20.8 \\
& &	(8)	&	(0.059)	&	(0.012)	&	(0.29)	&  (0.003)	&	(0.31)	&	(0.12)	&	&	\\
\\
B & 0\%--20\%	&	104	&	0.995	&	0.042	& 10.19	&	1.102	&	8.48	&	2.73	&	143.4/27=5.3	&	11.0/21=0.5 	& 	116.9/28=4.2 \\
& &	(5)	&	(0.055)	&	(0.004)	&	(0.46)	& (0.004)	&	(0.81)	&	(0.58)	&	&	\\
B & 20\%--60\%	&	113	&	0.905	&	0.067	& 7.44	&	1.182	&	4.95	&	2.86	&	206.5/27=7.6	& 39.4/21=1.9 	& 	95.7/28=3.4 \\
& &	(8)	&	(0.059)	&	(0.01)	&	(0.55)	& (0.008)	&	(0.91)	&	(0.52)	&	&	\\
 \end{tabular}\end{ruledtabular}
\label{table_bwpara}
\end{table*}

\subsubsection{Oscillation amplitudes with hydrodynamic models}
\label{sec:E2vsHydro}

The BW model~\cite{bw_hbt} suggests that the source eccentricity at 
freeze out is given by $\varepsilon_{\rm final} = 2R_{\rm s,2}^{2}/R_{\rm 
s,0}^{2} = -2R_{\rm o,2}^{2}/R_{\rm s,0}^{2} = 2R_{\rm os,2}^{2}/R_{\rm 
s,0}^{2}$ (see Eq.~\eqref{eq:cosine_series}) in the absence of 
position-momentum correlation, i.e. radial flow. In the presence of radial 
flow, the above relation would be smeared because the HBT radius does not 
reflect the whole source size, but the $\varepsilon_{\rm final}$ from 
$R_{\rm s,2}^{2}$ could still be a good estimator in the limit of $\kt=0$.

The HBT radii of both pions and kaon averaged over the azimuthal direction 
are well described by the hydrodynamic models including the BW model as 
shown so far. In this section, the oscillation amplitudes are also 
compared with the hydrodynamic models. The oscillation amplitudes were 
extracted by using Eq.~\eqref{eq:cosine_series}. The systematic uncertainties 
were estimated by performing the fitting with Eq.~\eqref{eq:cosine_series} 
for 
the data of various systematic sources described in Sec~\ref{sec:SE}. In 
Fig.~\ref{fig12} and Fig.\ref{fig13}, the oscillation amplitudes with 4 
different combinations of HBT radii are plotted in the form of a final 
eccentricity, $2R_{\rm \mu,2}^{2}/R_{\rm \nu,0}^{2}$, where $\mu$ and 
$\nu$ denote ${\rm o, s, os}$. The $R_{\rm \mu,2}^{2}$ is a fitting 
parameter in Eq.~\eqref{eq:cosine_series} that can take a negative value, 
which 
represents a different phase of the cosine function and shown in 
Fig.~\ref{fig12}(b,c) and Fig.~\ref{fig13}(b,c).

Figure~\ref{fig12}(a) shows the oscillation amplitude of $\Rs^{2}$ 
relative to the average, which is most sensitive to the final source 
eccentricity. The value of $2R_{\rm s,2}^{2}/R_{\rm s,0}^{2}$ increases 
with \mt, which would reflect \mt dependent ellipticity of the emission 
region. The other combinations of $|2R_{\rm \mu,2}^{2}/R_{\rm \nu,2}^{2}|$ 
show similar \mt dependence, but less dependence especially in 0\%--20\% 
centrality.  It should be noted that the \Ro and \Ros contain the particle 
emission duration in addition to the geometrical information, and they are 
also dominated by the anisotropy in the expansion velocity.

The data are compared with the blast-wave calculation using the Fit A 
parameters in Fig.~\ref{fig12}. The dependency on \mt of the oscillation 
amplitudes is not described well, although the \mt dependence of the mean 
radii is reproduced well. The large $\chi^{2}$ of HBT in Fit A in 
Table~\ref{table_bwpara} is mainly caused by such a discrepancy. The 
calculations from the event-by-event 3+1-D viscous-hydrodynamic 
model~\cite{Bozek} with a Glauber initial condition and shear viscosity 
$\eta$/s=0.08 (and also nonzero bulk viscosity) are compared to the same 
data in Fig.~\ref{fig13}. The model employs the equation of state with a 
crossover transition, but does not include a hadron cascade. The model 
quantitatively agrees with the data of \Rs and \Ros but overestimates \Ro 
in 20\%--60\% although the trend (not the magnitude) of dependency on \mt is 
reproduced in contrast with the blast-wave model.

When data points of kaons are compared to those of pions in the 0\%--20\% 
centrality sample, they mostly agree within the systematic uncertainties, 
but in the 20\%--60\% sample the $2R_{\rm s,2}^{2}/R_{\rm s,0}^{2}$($2R_{\rm 
o,2}^{2}/R_{\rm o,0}^{2}$) of kaons is slightly larger(smaller) than that 
of pions. More precise measurement is needed to confirm the difference.

\section{Summary and Conclusion\label{sec:summary}}

We have presented results from the PHENIX experiment on charged pion and 
kaon femtoscopy measurements in Au$+$Au collisions at $\sqsn$ = 200 GeV. In 
the azimuthal-integrated analysis, we have measured the HBT radii of both 
species with fine \mt and centrality bins. \mt scaling holds well for \Rs, 
but there are visible differences for \Ro and \Rl between charged pions 
and kaons at the same \mt, and the differences become larger in more 
central collisions. \mt scaling breaks for those radii, but \kt scaling 
works well for all radii. It is observed that the ratio $\Ro/\Rs$ of kaons 
is larger than that of pions, which may imply different emission 
durations. The hydrokinetic model was compared with our data. It 
reproduces most aspects of the data of both charged pions and kaons, but 
it fails to accurately describe the difference in \Ro.

In the azimuthal-dependent analysis, a first measurement of the HBT radii 
of charged kaons with respect to the second-order event plane has been 
performed and compared with pion measurements with finer \mt bins. 
Oscillation with respect to the event plane of kaon HBT radii has been 
clearly observed, and is similar to that of pions. The data were compared 
with the blast-wave model and the 3+1-D viscous-hydrodynamic model. The 
blast-wave model provides a good description of the overall trend of the 
\pt spectra, the elliptic flow, and the mean HBT radii, but fails to 
describe the details of femtoscopy measurements, such as the \mt dependent 
oscillation amplitude of the source radii.   While the 3+1-D 
viscous-hydrodynamic model does qualitatively reproduce the data, 
it overestimates the oscillation of \Ro.   We note that the viscous 
hydrodynamic model also reproduces well the other observables such as the 
\pt spectra and elliptic flow~\cite{Bozek_visH}.

Both the hydrokinetic model and viscous-hydrodynamic model surprisingly 
describe all aspects of the femtoscopic data, even though these models 
lack the shear viscosity of plasma and the microscopic transport phase. 
Including these effects may improve the description of the measured 
$\Ro/\Rs$ and $R_{\rm o,2}^{2}$.  More precise measurements and systematic 
model comparison for both azimuthal-dependent and azimuthal-integrated HBT 
measurements are needed.  The particle-species dependence, in addition to 
the differential femtoscopy measurements may help to elucidate the 
expansion dynamics of heavy ion collisions.

\section*{ACKNOWLEDGMENTS}   

We thank the staff of the Collider-Accelerator and Physics
Departments at Brookhaven National Laboratory and the staff of
the other PHENIX participating institutions for their vital
contributions.  We acknowledge support from the 
Office of Nuclear Physics in the
Office of Science of the Department of Energy,
the National Science Foundation, 
Abilene Christian University Research Council, 
Research Foundation of SUNY, and
Dean of the College of Arts and Sciences, Vanderbilt University 
(U.S.A),
Ministry of Education, Culture, Sports, Science, and Technology
and the Japan Society for the Promotion of Science (Japan),
Conselho Nacional de Desenvolvimento Cient\'{\i}fico e
Tecnol{\'o}gico and Funda\c c{\~a}o de Amparo {\`a} Pesquisa do
Estado de S{\~a}o Paulo (Brazil),
Natural Science Foundation of China (P.~R.~China),
Ministry of Science, Education, and Sports (Croatia),
Ministry of Education, Youth and Sports (Czech Republic),
Centre National de la Recherche Scientifique, Commissariat
{\`a} l'{\'E}nergie Atomique, and Institut National de Physique
Nucl{\'e}aire et de Physique des Particules (France),
Bundesministerium f\"ur Bildung und Forschung, Deutscher
Akademischer Austausch Dienst, and Alexander von Humboldt Stiftung (Germany),
National Science Fund, OTKA, K\'aroly R\'obert University College,
and the Ch. Simonyi Fund (Hungary),
Department of Atomic Energy and Department of Science and Technology (India), 
Israel Science Foundation (Israel), 
Basic Science Research Program through NRF of the Ministry of Education (Korea),
Physics Department, Lahore University of Management Sciences (Pakistan),
Ministry of Education and Science, Russian Academy of Sciences,
Federal Agency of Atomic Energy (Russia),
VR and Wallenberg Foundation (Sweden), 
the U.S. Civilian Research and Development Foundation for the
Independent States of the Former Soviet Union, 
the Hungarian American Enterprise Scholarship Fund,
and the US-Israel Binational Science Foundation.

\appendix

\section{Blast-wave model}

An emission function in the blast-wave parameterization for 
bosons~\cite{bw_hbt} is given by
\begin{align}
S(r, \phi_{s}, \tau, \eta) = m_{T} \cosh(\eta-y) \Omega(r,\phi_{s}) 
e^{-(\tau-\tau_{0})^{2}/(2\Delta\tau^{2})} \nonumber \\
\, \times \sum^{\infty}_{n=1} e^{n\alpha \cos(\phi_{b}-\phi_{p})} e^{-n\beta \cosh(\eta-y)},
\label{eq:Sfunc}
\end{align}
where $m_{T}$ is the transverse mass, 
$y$ is rapidity, $\phi_{p}$ is azimuthal angle of particle momentum, and 
$\alpha$ and $\beta$ are defined as
\begin{eqnarray}
\alpha &=& \frac{p_{T}}{T} \sinh \rho(r,\phi_{s}), \\
\beta &=& \frac{m_{T}}{T} \cosh \rho(r,\phi_{s}). 
\end{eqnarray}
The transverse rapidity $\rho(r,\phi_{s})$ is defined as:
\begin{eqnarray}
\rho(r,\phi_{s}) = (\rho_{0} + \rho_{2}\cos(2\phi_{b}) ) \tilde{r}, \nonumber \\
\tilde{r} = \sqrt{ (x/R_{\rm x})^{2} + (y/R_{\rm y})^{2} },
\end{eqnarray}
where $x$ and $y$ are space coordinate of particles, $\phi_{s}$ is 
azimuthal angle of the spatial positions, and $\phi_{b}$ is a boost 
direction. It is assumed that particles are boosted to the direction 
perpendicular to the elliptical subshell of the particle-emitting source, 
which satisfies the relation below:
\begin{equation}
\tan(\phi_{s}) = (R_{\rm y}/R_{\rm x})^{2} \tan(\phi_{b}).
\end{equation} 
The distribution of the source elements $\Omega(\tilde{r})$ is given by 
\begin{equation}
\Omega(\tilde{r}) = 1/( 1 + e^{(\tilde{r}-1)/a} ),
\end{equation} 
where $a$ denotes a surface diffuseness and $a=0$ gives a box profile and 
$a=0.3$ gives approximately a Gaussian profile.
Observables, such as spectra, $v_{2}$, and HBT radii, are obtained by 
performing the integral of the emission function Eq.~\eqref{eq:Sfunc} over 
phase space weighted with certain quantity $B$:
\begin{eqnarray}
\lefteqn{\int d^{4}x, S(x, K) B(x, K) = }\\ \nonumber 
&& \int_{0}^{2\pi} d\phi_{s} \int_{0}^{\infty} rdr \int_{-\infty}^{\infty} d\eta \int_{-\infty}^{\infty} \tau d\tau S(r, \phi_{s}, \tau, \eta) B(x, K).
\label{eq:obs}
\end{eqnarray}

Azimuthally integrated $p_{T}$ spectra can be obtained by integrating over 
$\phi_{p}$ and $\tau$ in Eq.~\eqref{eq:obs} setting $B(x,K)$=1. If we 
assume Boltzmann distribution for all particles, only the first term in 
the summation in Eq.~\eqref{eq:Sfunc} is used. Also, in case of analyzing 
particles in midrapidity region, Eq.~\eqref{eq:Sfunc} can be simplified 
by setting $y$=0. Then Eq.~\eqref{eq:obs} can be rewritten as the 
following:
\begin{eqnarray}
\frac{dN}{p_{T}dp_{T}} 
&=& \sqrt{2\pi} \, \tau_{0} \Delta\tau \, \int_{0}^{2\pi} d\phi_{p} \int_{0}^{2\pi} d\phi_{s} \int_{0}^{\infty} rdr \int_{-\infty}^{\infty} d\eta \nonumber 
\\
&\times& m_{T} \cosh(\eta) \Omega(r,\phi_{s}) e^{\alpha \cos(\phi_{b}-\phi_{p})} e^{-\beta \cosh(\eta)}, \nonumber \\
&=& 2\sqrt{2\pi} \, \tau_{0} \Delta\tau \, \int_{0}^{2\pi} d\phi_{p} \int_{0}^{2\pi} d\phi_{s} \int_{0}^{\infty} rdr \nonumber \\
&\times& m_{T} \Omega(r,\phi_{s}) \, e^{\alpha \cos(\phi_{b}-\phi_{p})} K_{1}(\beta),
\label{eq:spectra2}
\end{eqnarray}
where $K_{n}(\beta)$ is the modified Bessel function of the second kind, 
which is defined as
\begin{equation}
K_{n}(z) = \frac{1}{2} \int_{-\infty}^{\infty} dt \cosh(nt) e^{-z\cosh(t)}.
\end{equation}
Here we replace $\phi_{b}-\phi_{p}$ as $\phi^{\prime}$, and the range of 
the integral over $\phi^{\prime}$ is from $\phi_{b}$ to $\phi_{b}-2\pi$. 
Then the range can be replaced from 0 to $2\pi$ because the integrand is 
the periodic function with $2\pi$. Finally, Eq.~\eqref{eq:spectra2} is 
rewritten as
\begin{eqnarray}
\lefteqn{\frac{dN}{p_{T}dp_{T}} = 2(2\pi)^{3/2} \tau_{0} \Delta\tau m_{T}} \nonumber \\
&&\times \int_{0}^{2\pi} d\phi_{s} \int_{0}^{\infty} rdr \, \Omega(r,\phi_{s}) \, I_{0}(\alpha) K_{1}(\beta),
\label{eq:spectra3}
\end{eqnarray}
where $I_{n}$ is the modified Bessel function of the first kind given by
\begin{equation}
I_{n}(z) = \frac{1}{2\pi} \int_{0}^{2\pi} dt \cos(nt) e^{-z\cos(t)}.
\end{equation}

The elliptic flow $v_{2}$ is calculated as
\begin{equation}
v_{2}(p_{T},m) = \frac{ \int d\phi_{p} \int d^{4}x \, \cos(2\phi_{p}) S(x,K) }{ \int d\phi_{p} \int d^{4}x \, S(x,K) }.
\end{equation}
The denominator is the same expression with Eq.~\eqref{eq:spectra3}. The 
numerator can be calculated by a similar way to derive the $p_{T}$ 
spectra.

\begin{widetext}
\begin{eqnarray}
\lefteqn{\int d\phi_{p} \int d^{4}x \, \cos(2\phi_{p}) S(x,K),} \nonumber \\
&=& 2 \sqrt{2\pi} \, \tau_{0} \Delta\tau \int_{0}^{2\pi} d\phi_{p} \int_{0}^{2\pi} d\phi_{s} \int_{0}^{\infty} rdr 
\,\, m_{T} \Omega(r,\phi_{s}) \cos(2\phi_{p}) e^{\alpha \cos(\phi_{b}-\phi_{p})} K_{1}(\beta), \nonumber \\
&=& 2 \sqrt{2\pi} \, \tau_{0} \Delta\tau \int_{0}^{2\pi} d\phi_{p} \int_{0}^{\infty} rdr 
\, m_{T} \, \Omega(r,\phi_{s}) K_{1}(\beta) \cos(2\phi_{b}) \, \int_{0}^{2\pi} d\phi^{\prime} \cos(2\phi_{\prime}) e^{\alpha \cos(\phi^{\prime})}, \nonumber \\
&=& 2 (2\pi)^{3/2} \, \tau_{0} \Delta\tau m_{T} \int_{0}^{2\pi} d\phi_{p} \int_{0}^{\infty} rdr 
\,\, \Omega(r,\phi_{s}) K_{1}(\beta) \cos(2\phi_{b}) I_{2}(\alpha).
\end{eqnarray}
Finally, the elliptic flow can be expressed as
\begin{equation}
v_{2}(p_{T},m) = 
\frac{ \int_{0}^{2\pi} d\phi_{p} \int_{0}^{\infty} rdr \, \Omega(r,\phi_{s}) K_{1}(\beta) \cos(2\phi_{b}) I_{2}(\alpha) }
{\int_{0}^{2\pi} d\phi_{s} \int_{0}^{\infty} rdr \, \Omega(r,\phi_{s}) \, I_{0}(\alpha) K_{1}(\beta)}.
\end{equation}

The HBT radii are related to space-time variance as~\cite{bw_hbt}
\begin{eqnarray}
\Rs^{2} &=& \frac{1}{2}(\langle\tilde{x}^{2}\rangle+\langle\tilde{y}^{2}\rangle)-\frac{1}{2}(\langle\tilde{x}^{2}
\rangle-\langle\tilde{y}^{2}\rangle) \cos(2\phi_{p})-\langle\tilde{x}\tilde{y}\rangle \sin(2\phi_{p}), \\
\Ro^{2} &=& \frac{1}{2}(\langle\tilde{x}^{2}\rangle+\langle\tilde{y}^{2}\rangle)+\frac{1}{2}(\langle\tilde{x}^{2}
\rangle-\langle\tilde{y}^{2}\rangle) \cos(2\phi_{p})+\langle\tilde{x}\tilde{y}\rangle \sin(2\phi_{p}), \nonumber\\
& & -2\beta_{T}(\langle\tilde{t}\tilde{x}\rangle \cos\phi_{p}+\langle\tilde{t}\tilde{y}\rangle \sin\phi_{p})+\beta_{T}^{2}\langle\tilde{t}^{2}\rangle, \\
\Ros^{2} &=& \langle\tilde{x}\tilde{y}\rangle \cos(2\phi_{p})
-\frac{1}{2}(\langle\tilde{x}^{2}\rangle-\langle\tilde{y}^{2}\rangle) \sin(2\phi_{p})+\beta_{T}(\langle\tilde{t}\tilde{x}\rangle \sin\phi_{p}-\langle\tilde{t}\tilde{y}\rangle \cos\phi_{p}), \\
\Rl^{2} &=& \langle\tilde{z}^{2}\rangle 
-2\beta_{l}\langle\tilde{t}\tilde{z}\rangle+\beta_{l}^{2}\langle\tilde{t}^{2}\rangle, \nonumber \\
&=& \langle\tilde{z}^{2}\rangle,
\end{eqnarray}
\end{widetext}
where
\begin{eqnarray}
\langle f(x) \rangle &=& \frac{\int d^{4}x f(x) S(x, K)}{\int d^{4}x S(x, K)}, \\
\tilde{x}^{\mu} &=& x^{\mu} - \langle x^{\mu} \rangle,
\end{eqnarray}
and $\beta_{l}$ vanishes in the LCMS frame and the terms including $t$ and 
$z$ depend on the proper time $\tau$ and emission duration of particles 
$\Delta\tau$. As shown in above equations, \Rs depends on only the spatial 
extent of the source and azimuthal angle $\phi_{p}$, while \Ro and \Ros 
are also sensitive to $\tau$ and $\Delta\tau$ as well as the spatial 
extent.


\section{Data tables}

The extracted HBT radii and the oscillation amplitudes for charged pion 
and kaons in Au$+$Au collisions at $\sqsn$ =200 GeV are summarized in 
Table~\ref{data1}-\ref{data9}.

\begin{table*}[ht]
\caption{HBT parameters of positive pion pairs, shown as value $\pm$ 
statistical uncertainty [absolute value] $\pm$ systematic uncertainty [\%] 
for the centrality bins shown in Fig.~\protect\ref{fig3}.}
\begin{ruledtabular}\begin{tabular}{cccccc}
Centrality & $\langle\mt\rangle$ & $\lambda$ &     \Rs  &      \Ro  &      \Rl   \\ 
       &       [GeV/c]       &           &     [fm] &      [fm] &    [fm] \\ 
\hline
0\%--10\% & 0.3  &	  0.292 $\pm$ 0.003 $\pm$ 11.1 &	  5.11 $\pm$ 0.03 $\pm$ 0.5 &	  5.55 $\pm$ 0.04 $\pm$ 2.9 &	  6.3 $\pm$ 0.04 $\pm$ 2.3 \\
& 0.36  &	  0.358 $\pm$ 0.004 $\pm$ 8.7 &	  4.76 $\pm$ 0.03 $\pm$ 0.6 &	  5.22 $\pm$ 0.03 $\pm$ 3 &	  5.41 $\pm$ 0.04 $\pm$ 2.8 \\
& 0.41  &	  0.392 $\pm$ 0.004 $\pm$ 6.1 &	  4.52 $\pm$ 0.03 $\pm$ 0.8 &	  4.91 $\pm$ 0.03 $\pm$ 2.8 &	  4.8 $\pm$ 0.04 $\pm$ 3.7 \\
& 0.47  &	  0.402 $\pm$ 0.005 $\pm$ 11 &	  4.28 $\pm$ 0.03 $\pm$ 1.3 &	  4.51 $\pm$ 0.03 $\pm$ 4.7 &	  4.31 $\pm$ 0.03 $\pm$ 5.2 \\
& 0.53  &	  0.45 $\pm$ 0.006 $\pm$ 7.3 &	  4.17 $\pm$ 0.03 $\pm$ 0.7 &	  4.21 $\pm$ 0.04 $\pm$ 4.2 &	  4.08 $\pm$ 0.04 $\pm$ 4.2 \\
& 0.59  &	  0.454 $\pm$ 0.007 $\pm$ 5.9 &	  3.98 $\pm$ 0.03 $\pm$ 1.4 &	  3.97 $\pm$ 0.04 $\pm$ 5.7 &	  3.8 $\pm$ 0.04 $\pm$ 3.2 \\
& 0.66  &	  0.458 $\pm$ 0.007 $\pm$ 4.7 &	  3.77 $\pm$ 0.03 $\pm$ 1.1 &	  3.62 $\pm$ 0.04 $\pm$ 4.8 &	  3.45 $\pm$ 0.04 $\pm$ 4.8 \\
& 0.79  &	  0.462 $\pm$ 0.008 $\pm$ 6.1 &	  3.47 $\pm$ 0.03 $\pm$ 1.2 &	  3.27 $\pm$ 0.04 $\pm$ 4.9 &	  3.01 $\pm$ 0.04 $\pm$ 5.2 \\
& 0.96  &	  0.447 $\pm$ 0.016 $\pm$ 12 &	  3.16 $\pm$ 0.06 $\pm$ 3.1 &	  2.91 $\pm$ 0.07 $\pm$ 10.9 &	  2.49 $\pm$ 0.06 $\pm$ 6.1 \\
& 1.16  &	  0.448 $\pm$ 0.034 $\pm$ 18.5 &	  2.93 $\pm$ 0.11 $\pm$ 8.3 &	  2.4 $\pm$ 0.12 $\pm$ 14.4 &	  2.13 $\pm$ 0.11 $\pm$ 9.4 \\
 \\ 
10\%--20\%  & 0.3  &	  0.313 $\pm$ 0.003 $\pm$ 9 &	  4.63 $\pm$ 0.03 $\pm$ 0.6 &	  5.01 $\pm$ 0.03 $\pm$ 2.4 &	  5.77 $\pm$ 0.04 $\pm$ 1.9 \\
& 0.36  &	  0.379 $\pm$ 0.004 $\pm$ 7.2 &	  4.3 $\pm$ 0.03 $\pm$ 0.5 &	  4.74 $\pm$ 0.03 $\pm$ 2.6 &	  4.95 $\pm$ 0.04 $\pm$ 2.8 \\
& 0.41  &	  0.41 $\pm$ 0.004 $\pm$ 7.3 &	  4.1 $\pm$ 0.03 $\pm$ 1.3 &	  4.41 $\pm$ 0.03 $\pm$ 3.1 &	  4.44 $\pm$ 0.03 $\pm$ 3.5 \\
& 0.47  &	  0.441 $\pm$ 0.005 $\pm$ 7.5 &	  3.94 $\pm$ 0.03 $\pm$ 1 &	  4.12 $\pm$ 0.03 $\pm$ 3.6 &	  4.07 $\pm$ 0.03 $\pm$ 4.2 \\
& 0.53  &	  0.487 $\pm$ 0.006 $\pm$ 6.6 &	  3.83 $\pm$ 0.03 $\pm$ 0.8 &	  3.88 $\pm$ 0.03 $\pm$ 2.9 &	  3.83 $\pm$ 0.03 $\pm$ 3.7 \\
& 0.59  &	  0.501 $\pm$ 0.008 $\pm$ 5.5 &	  3.67 $\pm$ 0.03 $\pm$ 0.9 &	  3.68 $\pm$ 0.04 $\pm$ 3.7 &	  3.56 $\pm$ 0.04 $\pm$ 2.3 \\
& 0.66  &	  0.501 $\pm$ 0.008 $\pm$ 4.2 &	  3.5 $\pm$ 0.03 $\pm$ 1.5 &	  3.38 $\pm$ 0.03 $\pm$ 3.7 &	  3.24 $\pm$ 0.03 $\pm$ 3.5 \\
& 0.79  &	  0.501 $\pm$ 0.008 $\pm$ 4.3 &	  3.21 $\pm$ 0.03 $\pm$ 0.9 &	  2.99 $\pm$ 0.03 $\pm$ 4.9 &	  2.78 $\pm$ 0.03 $\pm$ 3.1 \\
& 0.96  &	  0.515 $\pm$ 0.016 $\pm$ 8.2 &	  2.94 $\pm$ 0.05 $\pm$ 1.4 &	  2.78 $\pm$ 0.06 $\pm$ 4.1 &	  2.34 $\pm$ 0.05 $\pm$ 6.3 \\
& 1.16  &	  0.52 $\pm$ 0.035 $\pm$ 11.2 &	  2.76 $\pm$ 0.1 $\pm$ 9 &	  2.23 $\pm$ 0.1 $\pm$ 2.8 &	  2.09 $\pm$ 0.1 $\pm$ 9.7 \\
 \\ 
20\%--40\%  & 0.3  &	  0.339 $\pm$ 0.003 $\pm$ 7.2 &	  4.09 $\pm$ 0.02 $\pm$ 0.6 &	  4.34 $\pm$ 0.02 $\pm$ 2.5 &	  5.01 $\pm$ 0.03 $\pm$ 1.5 \\
& 0.36  &	  0.401 $\pm$ 0.003 $\pm$ 5.8 &	  3.81 $\pm$ 0.02 $\pm$ 0.9 &	  4.09 $\pm$ 0.02 $\pm$ 2 &	  4.35 $\pm$ 0.03 $\pm$ 2.1 \\
& 0.41  &	  0.433 $\pm$ 0.004 $\pm$ 5.4 &	  3.63 $\pm$ 0.02 $\pm$ 0.7 &	  3.87 $\pm$ 0.02 $\pm$ 2.3 &	  3.88 $\pm$ 0.03 $\pm$ 2.2 \\
& 0.47  &	  0.46 $\pm$ 0.005 $\pm$ 6.9 &	  3.5 $\pm$ 0.02 $\pm$ 0.9 &	  3.59 $\pm$ 0.02 $\pm$ 3.2 &	  3.55 $\pm$ 0.03 $\pm$ 3.7 \\
& 0.53  &	  0.493 $\pm$ 0.005 $\pm$ 5.8 &	  3.39 $\pm$ 0.02 $\pm$ 0.8 &	  3.38 $\pm$ 0.02 $\pm$ 2.9 &	  3.3 $\pm$ 0.03 $\pm$ 3 \\
& 0.59  &	  0.519 $\pm$ 0.007 $\pm$ 3.7 &	  3.26 $\pm$ 0.02 $\pm$ 0.3 &	  3.2 $\pm$ 0.03 $\pm$ 2.7 &	  3.11 $\pm$ 0.03 $\pm$ 2.1 \\
& 0.66  &	  0.505 $\pm$ 0.006 $\pm$ 3.2 &	  3.07 $\pm$ 0.02 $\pm$ 0.4 &	  2.91 $\pm$ 0.02 $\pm$ 2.7 &	  2.77 $\pm$ 0.03 $\pm$ 1.9 \\
& 0.79  &	  0.532 $\pm$ 0.007 $\pm$ 3.5 &	  2.89 $\pm$ 0.02 $\pm$ 1.1 &	  2.66 $\pm$ 0.02 $\pm$ 3.7 &	  2.45 $\pm$ 0.02 $\pm$ 3.1 \\
& 0.96  &	  0.54 $\pm$ 0.014 $\pm$ 7.7 &	  2.63 $\pm$ 0.04 $\pm$ 4.3 &	  2.34 $\pm$ 0.04 $\pm$ 3.5 &	  2.07 $\pm$ 0.04 $\pm$ 3.2 \\
& 1.16  &	  0.554 $\pm$ 0.027 $\pm$ 4.3 &	  2.35 $\pm$ 0.08 $\pm$ 2.9 &	  2.09 $\pm$ 0.08 $\pm$ 5.5 &	  1.76 $\pm$ 0.07 $\pm$ 5.6 \\
 \\ 
40\%--70\%  & 0.3  &	  0.363 $\pm$ 0.004 $\pm$ 5.8 &	  3.27 $\pm$ 0.02 $\pm$ 0.9 &	  3.39 $\pm$ 0.03 $\pm$ 2.6 &	  3.94 $\pm$ 0.03 $\pm$ 1.1 \\
& 0.36  &	  0.426 $\pm$ 0.005 $\pm$ 2.8 &	  3.05 $\pm$ 0.03 $\pm$ 0.6 &	  3.29 $\pm$ 0.03 $\pm$ 1.7 &	  3.42 $\pm$ 0.03 $\pm$ 1.1 \\
& 0.41  &	  0.455 $\pm$ 0.006 $\pm$ 4.2 &	  2.92 $\pm$ 0.03 $\pm$ 0.7 &	  3.16 $\pm$ 0.03 $\pm$ 3 &	  3.09 $\pm$ 0.03 $\pm$ 1.5 \\
& 0.47  &	  0.493 $\pm$ 0.007 $\pm$ 3.6 &	  2.87 $\pm$ 0.03 $\pm$ 1.2 &	  2.94 $\pm$ 0.03 $\pm$ 1.7 &	  2.89 $\pm$ 0.03 $\pm$ 2.1 \\
& 0.53  &	  0.509 $\pm$ 0.008 $\pm$ 3.9 &	  2.73 $\pm$ 0.03 $\pm$ 0.9 &	  2.81 $\pm$ 0.03 $\pm$ 1.8 &	  2.63 $\pm$ 0.03 $\pm$ 2 \\
& 0.59  &	  0.54 $\pm$ 0.01 $\pm$ 3.4 &	  2.7 $\pm$ 0.03 $\pm$ 1.2 &	  2.67 $\pm$ 0.04 $\pm$ 1.7 &	  2.46 $\pm$ 0.04 $\pm$ 1.6 \\
& 0.66  &	  0.569 $\pm$ 0.01 $\pm$ 1.2 &	  2.61 $\pm$ 0.03 $\pm$ 2.1 &	  2.48 $\pm$ 0.03 $\pm$ 1.9 &	  2.31 $\pm$ 0.03 $\pm$ 1 \\
& 0.79  &	  0.551 $\pm$ 0.01 $\pm$ 4.6 &	  2.36 $\pm$ 0.03 $\pm$ 1.6 &	  2.1 $\pm$ 0.03 $\pm$ 2.1 &	  1.93 $\pm$ 0.03 $\pm$ 2.1 \\
& 0.96  &	  0.619 $\pm$ 0.024 $\pm$ 5.7 &	  2.35 $\pm$ 0.06 $\pm$ 1 &	  1.99 $\pm$ 0.06 $\pm$ 4.4 &	  1.82 $\pm$ 0.06 $\pm$ 1.6 \\
& 1.16  &	  0.592 $\pm$ 0.041 $\pm$ 3.8 &	  2.02 $\pm$ 0.13 $\pm$ 2.8 &	  1.64 $\pm$ 0.13 $\pm$ 11.1 &	  1.45 $\pm$ 0.11 $\pm$ 3.5 \\
 \end{tabular}\end{ruledtabular}
\label{data1}
\end{table*}

\begin{table*}[ht]
\caption{HBT parameters of negative pion pairs, shown as value $\pm$ 
statistical uncertainty [absolute value] $\pm$ systematic uncertainty [\%] 
for the centrality bins shown in Fig.~\protect\ref{fig3}.}
\begin{ruledtabular}\begin{tabular}{cccccc}
Centrality & $\langle\mt\rangle$ & $\lambda$ &     \Rs  &      \Ro  &      \Rl \\ 
           &       [GeV/c]       &           &     [fm] &      [fm] &      [fm] \\ 
\hline
0\%--10\% & 0.3  &	  0.275 $\pm$ 0.003 $\pm$ 10.5 &	  5.11 $\pm$ 0.03 $\pm$ 0.6 &	  5.49 $\pm$ 0.04 $\pm$ 3 &	  6.28 $\pm$ 0.05 $\pm$ 2.9 \\
& 0.36  &	  0.353 $\pm$ 0.004 $\pm$ 7.3 &	  4.74 $\pm$ 0.03 $\pm$ 0.7 &	  5.25 $\pm$ 0.04 $\pm$ 2.1 &	  5.44 $\pm$ 0.05 $\pm$ 3.4 \\
& 0.41  &	  0.387 $\pm$ 0.005 $\pm$ 7.9 &	  4.5 $\pm$ 0.03 $\pm$ 0.4 &	  4.91 $\pm$ 0.04 $\pm$ 3.6 &	  4.82 $\pm$ 0.04 $\pm$ 3.9 \\
& 0.47  &	  0.399 $\pm$ 0.005 $\pm$ 9.1 &	  4.31 $\pm$ 0.03 $\pm$ 0.9 &	  4.38 $\pm$ 0.04 $\pm$ 4.3 &	  4.29 $\pm$ 0.04 $\pm$ 5.4 \\
& 0.53  &	  0.444 $\pm$ 0.007 $\pm$ 4.7 &	  4.17 $\pm$ 0.03 $\pm$ 1.2 &	  4.19 $\pm$ 0.04 $\pm$ 3.6 &	  4.11 $\pm$ 0.04 $\pm$ 3 \\
& 0.59  &	  0.45 $\pm$ 0.008 $\pm$ 7.4 &	  3.95 $\pm$ 0.04 $\pm$ 1.2 &	  3.91 $\pm$ 0.04 $\pm$ 5.7 &	  3.82 $\pm$ 0.04 $\pm$ 3.8 \\
& 0.66  &	  0.451 $\pm$ 0.008 $\pm$ 7.3 &	  3.77 $\pm$ 0.03 $\pm$ 0.2 &	  3.61 $\pm$ 0.04 $\pm$ 5.3 &	  3.47 $\pm$ 0.04 $\pm$ 3.3 \\
& 0.79  &	  0.442 $\pm$ 0.008 $\pm$ 6.2 &	  3.49 $\pm$ 0.03 $\pm$ 1 &	  3.16 $\pm$ 0.04 $\pm$ 6 &	  2.96 $\pm$ 0.04 $\pm$ 5.5 \\
& 0.96  &	  0.437 $\pm$ 0.017 $\pm$ 5 &	  3.22 $\pm$ 0.07 $\pm$ 2.4 &	  2.78 $\pm$ 0.07 $\pm$ 4.5 &	  2.46 $\pm$ 0.06 $\pm$ 6.9 \\
& 1.15  &	  0.526 $\pm$ 0.046 $\pm$ 13.3 &	  3.1 $\pm$ 0.13 $\pm$ 3.1 &	  2.27 $\pm$ 0.13 $\pm$ 8.1 &	  2.43 $\pm$ 0.13 $\pm$ 11.1 \\
\\
10\%--20\% & 0.3  &	  0.295 $\pm$ 0.003 $\pm$ 9.4 &	  4.63 $\pm$ 0.03 $\pm$ 0.4 &	  4.93 $\pm$ 0.04 $\pm$ 3.4 &	  5.74 $\pm$ 0.04 $\pm$ 2.3 \\
& 0.36  &	  0.363 $\pm$ 0.004 $\pm$ 6.4 &	  4.3 $\pm$ 0.03 $\pm$ 0.3 &	  4.71 $\pm$ 0.04 $\pm$ 3.1 &	  4.88 $\pm$ 0.04 $\pm$ 2.6 \\
& 0.41  &	  0.404 $\pm$ 0.005 $\pm$ 6.9 &	  4.13 $\pm$ 0.03 $\pm$ 0.5 &	  4.42 $\pm$ 0.04 $\pm$ 2.8 &	  4.46 $\pm$ 0.04 $\pm$ 3.6 \\
& 0.47  &	  0.443 $\pm$ 0.006 $\pm$ 7.2 &	  3.95 $\pm$ 0.03 $\pm$ 1 &	  4.13 $\pm$ 0.03 $\pm$ 2.9 &	  4.05 $\pm$ 0.04 $\pm$ 4.1 \\
& 0.53  &	  0.462 $\pm$ 0.007 $\pm$ 5.9 &	  3.8 $\pm$ 0.03 $\pm$ 0.2 &	  3.91 $\pm$ 0.04 $\pm$ 3.4 &	  3.77 $\pm$ 0.04 $\pm$ 2.8 \\
& 0.59  &	  0.492 $\pm$ 0.008 $\pm$ 4.1 &	  3.68 $\pm$ 0.03 $\pm$ 0.4 &	  3.64 $\pm$ 0.04 $\pm$ 3.7 &	  3.53 $\pm$ 0.04 $\pm$ 3.5 \\
& 0.66  &	  0.48 $\pm$ 0.008 $\pm$ 4.7 &	  3.47 $\pm$ 0.03 $\pm$ 1.2 &	  3.33 $\pm$ 0.03 $\pm$ 4.3 &	  3.15 $\pm$ 0.03 $\pm$ 3.8 \\
& 0.79  &	  0.497 $\pm$ 0.009 $\pm$ 5.1 &	  3.25 $\pm$ 0.03 $\pm$ 0.5 &	  3.01 $\pm$ 0.03 $\pm$ 4 &	  2.75 $\pm$ 0.03 $\pm$ 3.7 \\
& 0.96  &	  0.482 $\pm$ 0.016 $\pm$ 3.6 &	  2.89 $\pm$ 0.05 $\pm$ 1.2 &	  2.68 $\pm$ 0.06 $\pm$ 4.5 &	  2.27 $\pm$ 0.05 $\pm$ 5.2 \\
& 1.15  &	  0.579 $\pm$ 0.043 $\pm$ 16.5 &	  2.67 $\pm$ 0.12 $\pm$ 8.6 &	  2.29 $\pm$ 0.11 $\pm$ 5.7 &	  2.11 $\pm$ 0.1 $\pm$ 6.7 \\
\\
20\%--40\% & 0.3  &	  0.32 $\pm$ 0.003 $\pm$ 6.8 &	  4.11 $\pm$ 0.02 $\pm$ 0.5 &	  4.28 $\pm$ 0.03 $\pm$ 2.3 &	  5.05 $\pm$ 0.03 $\pm$ 2 \\
& 0.36  &	  0.385 $\pm$ 0.004 $\pm$ 7.3 &	  3.83 $\pm$ 0.03 $\pm$ 1.4 &	  4.03 $\pm$ 0.03 $\pm$ 2.8 &	  4.32 $\pm$ 0.03 $\pm$ 2.1 \\
& 0.41  &	  0.429 $\pm$ 0.005 $\pm$ 5 &	  3.65 $\pm$ 0.02 $\pm$ 0.5 &	  3.89 $\pm$ 0.03 $\pm$ 1.3 &	  3.93 $\pm$ 0.03 $\pm$ 2.5 \\
& 0.47  &	  0.458 $\pm$ 0.005 $\pm$ 6.6 &	  3.5 $\pm$ 0.02 $\pm$ 0.5 &	  3.59 $\pm$ 0.03 $\pm$ 3.1 &	  3.54 $\pm$ 0.03 $\pm$ 3.5 \\
& 0.53  &	  0.487 $\pm$ 0.006 $\pm$ 4.5 &	  3.4 $\pm$ 0.03 $\pm$ 0.7 &	  3.38 $\pm$ 0.03 $\pm$ 1.8 &	  3.32 $\pm$ 0.03 $\pm$ 2.7 \\
& 0.59  &	  0.509 $\pm$ 0.007 $\pm$ 3.5 &	  3.26 $\pm$ 0.03 $\pm$ 0.9 &	  3.18 $\pm$ 0.03 $\pm$ 2.7 &	  3.1 $\pm$ 0.03 $\pm$ 2.9 \\
& 0.66  &	  0.521 $\pm$ 0.007 $\pm$ 2.5 &	  3.11 $\pm$ 0.02 $\pm$ 0.6 &	  2.96 $\pm$ 0.03 $\pm$ 2.6 &	  2.86 $\pm$ 0.03 $\pm$ 2.2 \\
& 0.79  &	  0.521 $\pm$ 0.007 $\pm$ 4.8 &	  2.87 $\pm$ 0.02 $\pm$ 0.7 &	  2.62 $\pm$ 0.03 $\pm$ 3 &	  2.4 $\pm$ 0.02 $\pm$ 3 \\
& 0.96  &	  0.536 $\pm$ 0.015 $\pm$ 2.2 &	  2.66 $\pm$ 0.04 $\pm$ 1.1 &	  2.31 $\pm$ 0.04 $\pm$ 3 &	  2.05 $\pm$ 0.04 $\pm$ 3.3 \\
& 1.16  &	  0.565 $\pm$ 0.034 $\pm$ 8.6 &	  2.45 $\pm$ 0.1 $\pm$ 3.4 &	  1.97 $\pm$ 0.09 $\pm$ 1.7 &	  1.81 $\pm$ 0.09 $\pm$ 3.3 \\
\\
40\%--70\% & 0.3  &	  0.344 $\pm$ 0.004 $\pm$ 5.2 &	  3.25 $\pm$ 0.03 $\pm$ 0.9 &	  3.31 $\pm$ 0.03 $\pm$ 1.5 &	  3.96 $\pm$ 0.04 $\pm$ 0.7 \\
& 0.36  &	  0.409 $\pm$ 0.006 $\pm$ 4.3 &	  3.05 $\pm$ 0.03 $\pm$ 0.9 &	  3.27 $\pm$ 0.03 $\pm$ 1.9 &	  3.41 $\pm$ 0.04 $\pm$ 1.8 \\
& 0.41  &	  0.452 $\pm$ 0.007 $\pm$ 3.1 &	  2.92 $\pm$ 0.03 $\pm$ 1 &	  3.15 $\pm$ 0.03 $\pm$ 2.3 &	  3.1 $\pm$ 0.04 $\pm$ 1.3 \\
& 0.47  &	  0.489 $\pm$ 0.008 $\pm$ 4.8 &	  2.83 $\pm$ 0.03 $\pm$ 1.4 &	  2.95 $\pm$ 0.03 $\pm$ 2.6 &	  2.86 $\pm$ 0.04 $\pm$ 1.8 \\
& 0.53  &	  0.528 $\pm$ 0.009 $\pm$ 2.5 &	  2.8 $\pm$ 0.03 $\pm$ 1.1 &	  2.77 $\pm$ 0.03 $\pm$ 2 &	  2.67 $\pm$ 0.04 $\pm$ 2.2 \\
& 0.59  &	  0.539 $\pm$ 0.011 $\pm$ 2.9 &	  2.65 $\pm$ 0.04 $\pm$ 2.3 &	  2.62 $\pm$ 0.04 $\pm$ 3.2 &	  2.47 $\pm$ 0.04 $\pm$ 1.3 \\
& 0.66  &	  0.532 $\pm$ 0.01 $\pm$ 4.6 &	  2.5 $\pm$ 0.03 $\pm$ 2 &	  2.4 $\pm$ 0.03 $\pm$ 2.6 &	  2.25 $\pm$ 0.03 $\pm$ 3.7 \\
& 0.79  &	  0.565 $\pm$ 0.012 $\pm$ 3.5 &	  2.41 $\pm$ 0.03 $\pm$ 1.4 &	  2.19 $\pm$ 0.04 $\pm$ 1.7 &	  1.99 $\pm$ 0.03 $\pm$ 2.5 \\
& 0.96  &	  0.582 $\pm$ 0.023 $\pm$ 7.7 &	  2.26 $\pm$ 0.07 $\pm$ 2.2 &	  1.91 $\pm$ 0.07 $\pm$ 3.9 &	  1.7 $\pm$ 0.06 $\pm$ 3.5 \\
& 1.15  &	  0.641 $\pm$ 0.063 $\pm$ 3.5 &	  2.18 $\pm$ 0.17 $\pm$ 9 &	  1.72 $\pm$ 0.14 $\pm$ 10.9 &	  1.67 $\pm$ 0.15 $\pm$ 17.3 \\
 \end{tabular}\end{ruledtabular}
\label{data2}
\end{table*}

\begin{table*}[ht]
\caption{HBT parameters of charge-combined kaon pairs, shown as value $\pm$ 
statistical uncertainty [absolute value] $\pm$ systematic uncertainty [\%] 
for the centrality bins shown in Fig.~\protect\ref{fig3}.}
\begin{ruledtabular}\begin{tabular}{cccccc}
Centrality & $\langle\mt\rangle$ & $\lambda$ &     \Rs  &      \Ro  &      \Rl \\ 
           &       [GeV/c]       &           &     [fm] &      [fm] &      [fm] \\ 
\hline
0\%--10\% & 0.76  &	  0.507 $\pm$ 0.044 $\pm$ 12.9 &	  3.8 $\pm$ 0.15 $\pm$ 2.7 &	  4.07 $\pm$ 0.15 $\pm$ 5 &	  4.09 $\pm$ 0.17 $\pm$ 2.9 \\
& 0.91  &	  0.46 $\pm$ 0.043 $\pm$ 12.7 &	  3.32 $\pm$ 0.13 $\pm$ 4.9 &	  3.49 $\pm$ 0.15 $\pm$ 5.3 &	  3.25 $\pm$ 0.17 $\pm$ 4.1 \\
& 1.1  &	  0.404 $\pm$ 0.043 $\pm$ 4.3 &	  2.86 $\pm$ 0.14 $\pm$ 3.4 &	  3.1 $\pm$ 0.18 $\pm$ 5.4 &	  2.36 $\pm$ 0.15 $\pm$ 7.7 \\
 \\
10\%--20\% & 0.76  &	  0.539 $\pm$ 0.044 $\pm$ 11.3 &	  3.37 $\pm$ 0.13 $\pm$ 2.6 &	  3.56 $\pm$ 0.13 $\pm$ 4 &	  3.55 $\pm$ 0.16 $\pm$ 4.5 \\
& 0.91  &	  0.515 $\pm$ 0.044 $\pm$ 15.3 &	  3.03 $\pm$ 0.12 $\pm$ 2.7 &	  3.2 $\pm$ 0.13 $\pm$ 1.2 &	  2.98 $\pm$ 0.14 $\pm$ 12.5 \\
& 1.1  &	  0.457 $\pm$ 0.052 $\pm$ 14.7 &	  2.75 $\pm$ 0.14 $\pm$ 1.5 &	  2.48 $\pm$ 0.14 $\pm$ 2.6 &	  2.34 $\pm$ 0.17 $\pm$ 10.8 \\
 \\
20\%--40\% & 0.76  &	  0.64 $\pm$ 0.044 $\pm$ 11.4 &	  3.18 $\pm$ 0.11 $\pm$ 2 &	  3.28 $\pm$ 0.1 $\pm$ 4 &	  3.15 $\pm$ 0.12 $\pm$ 4.7 \\
& 0.91  &	  0.489 $\pm$ 0.034 $\pm$ 7.6 &	  2.69 $\pm$ 0.1 $\pm$ 2.8 &	  2.73 $\pm$ 0.1 $\pm$ 3 &	  2.29 $\pm$ 0.11 $\pm$ 3.8 \\
& 1.09  &	  0.501 $\pm$ 0.046 $\pm$ 13.9 &	  2.56 $\pm$ 0.13 $\pm$ 3.8 &	  2.31 $\pm$ 0.12 $\pm$ 3.2 &	  2.05 $\pm$ 0.13 $\pm$ 7.4 \\
 \\
40\%--70\% & 0.76  &	  0.624 $\pm$ 0.062 $\pm$ 5.1 &	  2.73 $\pm$ 0.16 $\pm$ 5.3 &	  2.67 $\pm$ 0.16 $\pm$ 2.5 &	  2.35 $\pm$ 0.15 $\pm$ 8.3 \\
& 0.91  &	  0.565 $\pm$ 0.065 $\pm$ 14.6 &	  2.33 $\pm$ 0.17 $\pm$ 9.8 &	  2.44 $\pm$ 0.16 $\pm$ 5.5 &	  1.97 $\pm$ 0.17 $\pm$ 15.4 \\
& 1.09  &	  0.575 $\pm$ 0.068 $\pm$ 19.5 &	  2.08 $\pm$ 0.17 $\pm$ 5.8 &	  2.03 $\pm$ 0.19 $\pm$ 5.3 &	  1.54 $\pm$ 0.16 $\pm$ 8 \\
 \end{tabular}\end{ruledtabular}
\label{data3}
\end{table*}

\begin{table*}[ht]
\caption{\Ro/\Rs of positive and negative pion pairs plus 
charge-combined kaon pairs, shown as value $\pm$ 
statistical uncertainty [absolute value] $\pm$ systematic uncertainty [\%] 
for the centrality bins shown in Fig.~\protect\ref{fig4}.}
\begin{ruledtabular}\begin{tabular}{cccccc}
meson  & $\langle\mt\rangle$ &	 \multicolumn{4}{c}{\Ro/\Rs} \\ 
pair  &   [GeV/c] &	 0\%--10\% &	 10\%--20\% &	 20\%--40\% &	 40\%--70\% \\ \hline
 $\pi^{+}\pi^{+}$ & 0.3   &	  1.09 $\pm$ 0.01 $\pm$ 3.3 &	  1.08 $\pm$ 0.01 $\pm$ 2.3 &	  1.06 $\pm$ 0.01 $\pm$ 2.1 &	  1.04 $\pm$ 0.01 $\pm$ 1.9 \\
& 0.36  &	  1.10 $\pm$ 0.01 $\pm$ 3.5 &	  1.10 $\pm$ 0.01 $\pm$ 3.0 &	  1.07 $\pm$ 0.01 $\pm$ 1.6 &	  1.08 $\pm$ 0.01 $\pm$ 1.7 \\
& 0.41  &	  1.09 $\pm$ 0.01 $\pm$ 3.5 &	  1.08 $\pm$ 0.01 $\pm$ 3.0 &	  1.07 $\pm$ 0.01 $\pm$ 2.1 &	  1.08 $\pm$ 0.01 $\pm$ 2.4 \\
& 0.47  &	  1.05 $\pm$ 0.01 $\pm$ 4.3 &	  1.04 $\pm$ 0.01 $\pm$ 3.6 &	  1.03 $\pm$ 0.01 $\pm$ 2.5 &	  1.03 $\pm$ 0.01 $\pm$ 0.9 \\
& 0.53  &	  1.01 $\pm$ 0.01 $\pm$ 3.8 &	  1.01 $\pm$ 0.01 $\pm$ 2.3 &	  1.00 $\pm$ 0.01 $\pm$ 2.2 &	  1.03 $\pm$ 0.02 $\pm$ 1.1 \\
& 0.59  &	  1.00 $\pm$ 0.01 $\pm$ 5.6 &	  1.00 $\pm$ 0.01 $\pm$ 3.7 &	  0.98 $\pm$ 0.01 $\pm$ 2.6 &	  0.99 $\pm$ 0.02 $\pm$ 1.4 \\
& 0.66  &	  0.96 $\pm$ 0.01 $\pm$ 5.6 &	  0.96 $\pm$ 0.01 $\pm$ 4.7 &	  0.95 $\pm$ 0.01 $\pm$ 2.9 &	  0.95 $\pm$ 0.02 $\pm$ 3.2 \\
& 0.79  &	  0.94 $\pm$ 0.01 $\pm$ 5.4 &	  0.93 $\pm$ 0.01 $\pm$ 5.5 &	  0.92 $\pm$ 0.01 $\pm$ 4.7 &	  0.89 $\pm$ 0.02 $\pm$ 1.5 \\
& 0.96  &	  0.92 $\pm$ 0.03 $\pm$ 10.7 &	  0.94 $\pm$ 0.03 $\pm$ 5.0 &	  0.89 $\pm$ 0.02 $\pm$ 4.5 &	  0.85 $\pm$ 0.03 $\pm$ 3.7 \\
& 1.16  &	  0.82 $\pm$ 0.05 $\pm$ 7.8 &	  0.81 $\pm$ 0.05 $\pm$ 11.9 &	  0.89 $\pm$ 0.04 $\pm$ 2.9 &	  0.81 $\pm$ 0.08 $\pm$ 10.5 \\
 \\
$\pi^{-}\pi^{-}$ & 0.3   &	  1.07 $\pm$ 0.01 $\pm$ 3.3 &	  1.07 $\pm$ 0.01 $\pm$ 3.2 &	  1.04 $\pm$ 0.01 $\pm$ 2.1 &	  1.02 $\pm$ 0.01 $\pm$ 1.5 \\
& 0.36  &	  1.11 $\pm$ 0.01 $\pm$ 2.5 &	  1.10 $\pm$ 0.01 $\pm$ 3.2 &	  1.05 $\pm$ 0.01 $\pm$ 1.9 &	  1.07 $\pm$ 0.02 $\pm$ 2.4 \\
& 0.41  &	  1.09 $\pm$ 0.01 $\pm$ 3.9 &	  1.07 $\pm$ 0.01 $\pm$ 3.1 &	  1.07 $\pm$ 0.01 $\pm$ 1.2 &	  1.08 $\pm$ 0.02 $\pm$ 1.6 \\
& 0.47  &	  1.02 $\pm$ 0.01 $\pm$ 4.3 &	  1.04 $\pm$ 0.01 $\pm$ 3.2 &	  1.03 $\pm$ 0.01 $\pm$ 2.7 &	  1.04 $\pm$ 0.02 $\pm$ 2.1 \\
& 0.53  &	  1.01 $\pm$ 0.01 $\pm$ 4.3 &	  1.03 $\pm$ 0.01 $\pm$ 3.4 &	  0.99 $\pm$ 0.01 $\pm$ 1.3 &	  0.99 $\pm$ 0.02 $\pm$ 1.6 \\
& 0.59  &	  0.99 $\pm$ 0.01 $\pm$ 5.4 &	  0.99 $\pm$ 0.01 $\pm$ 3.9 &	  0.98 $\pm$ 0.01 $\pm$ 3.0 &	  0.99 $\pm$ 0.02 $\pm$ 4.4 \\
& 0.66  &	  0.96 $\pm$ 0.01 $\pm$ 5.5 &	  0.96 $\pm$ 0.01 $\pm$ 4.9 &	  0.95 $\pm$ 0.01 $\pm$ 3.2 &	  0.96 $\pm$ 0.02 $\pm$ 1.9 \\
& 0.79  &	  0.91 $\pm$ 0.01 $\pm$ 6.7 &	  0.92 $\pm$ 0.01 $\pm$ 4.3 &	  0.91 $\pm$ 0.01 $\pm$ 2.9 &	  0.91 $\pm$ 0.02 $\pm$ 2.0 \\
& 0.96  &	  0.86 $\pm$ 0.03 $\pm$ 5.0 &	  0.93 $\pm$ 0.03 $\pm$ 4.3 &	  0.87 $\pm$ 0.02 $\pm$ 3.8 &	  0.85 $\pm$ 0.04 $\pm$ 3.3 \\
& 1.15  &	  0.73 $\pm$ 0.05 $\pm$ 8.7 &	  0.86 $\pm$ 0.06 $\pm$ 7.6 &	  0.8 $\pm$ 0.05 $\pm$ 4.8 &	  0.79 $\pm$ 0.09 $\pm$ 6.7 \\
\\
$K^{+}K^{+}+K^{-}K^{-}$ & 0.76  &	  1.07 $\pm$ 0.06 $\pm$ 5.3 &	  1.06 $\pm$ 0.06 $\pm$ 4.5 &	  1.03 $\pm$ 0.05 $\pm$ 5.3 &	  0.98 $\pm$ 0.08 $\pm$ 7.1 \\
& 0.91  &	  1.05 $\pm$ 0.06 $\pm$ 8.9 &	  1.06 $\pm$ 0.06 $\pm$ 2.9 &	  1.01 $\pm$ 0.05 $\pm$ 3.6 &	  1.05 $\pm$ 0.10 $\pm$ 6.1 \\
& 1.1   &	  1.08 $\pm$ 0.08 $\pm$ 5.2 &	  0.90 $\pm$ 0.07 $\pm$ 3.8 &	  0.90 $\pm$ 0.07 $\pm$ 3.1 &	  0.98 $\pm$ 0.12 $\pm$ 2.3 \\
 \end{tabular}\end{ruledtabular}
\label{data45}
\end{table*}



\begin{table*}[ht]
\caption{Azimuthal angle dependence of HBT radii of charged pions, 
shown as value $\pm$ statistical uncertainty [absolute value] 
$\pm$ systematic uncertainty [\%] for the 0\%--20\% and 
20\%--60\% centrality bins.}
\begin{ruledtabular}\begin{tabular}{ccccccc}
Centrality & \kt      & $\phi-\Psi_2$  & $\Rs^2$  &	$\Ro^2$  & $\Rl^2$  & $\Ros^2$ \\ 
           & [GeV/c]  & [rad]          & [fm$^2$] &	[fm$^2$] & [fm$^2$] & [fm$^2$] \\ 
\hline
0\%--20\% & 0.2--0.3 &$0$	 & 23.68 $\pm$ 0.27 $\pm$ 0.56	 & 25.25 $\pm$ 0.34 $\pm$ 1.54	 & 34.72 $\pm$ 0.48 $\pm$ 1.97	 & 0.21 $\pm$ 0.2 $\pm$ 0.11	 \\ 
&&$\pi/4$	 & 22.88 $\pm$ 0.27 $\pm$ 0.56	 & 27.33 $\pm$ 0.37 $\pm$ 1.79	 & 35.05 $\pm$ 0.49 $\pm$ 1.63	 & 2.53 $\pm$ 0.21 $\pm$ 0.25	 \\ 
&&$\pi/2$	 & 22.46 $\pm$ 0.27 $\pm$ 0.56	 & 30.65 $\pm$ 0.42 $\pm$ 1.8	 & 35.34 $\pm$ 0.51 $\pm$ 0.98	 & 0.26 $\pm$ 0.23 $\pm$ 0.27	 \\ 
&&$3\pi/4$	 & 23.4 $\pm$ 0.28 $\pm$ 0.51	 & 28.06 $\pm$ 0.39 $\pm$ 1.31	 & 35.46 $\pm$ 0.51 $\pm$ 1.41	 & -2.01 $\pm$ 0.22 $\pm$ 0.24	 \\ 
\\
&0.3--0.4 &$0$	 & 20.5 $\pm$ 0.2 $\pm$ 0.67	 & 21.87 $\pm$ 0.24 $\pm$ 1.38	 & 24.44 $\pm$ 0.29 $\pm$ 1.64	 & 0.22 $\pm$ 0.14 $\pm$ 0.13	 \\ 
&&$\pi/4$	 & 19.82 $\pm$ 0.21 $\pm$ 0.4	 & 24.09 $\pm$ 0.27 $\pm$ 1.2	 & 24.03 $\pm$ 0.3 $\pm$ 1.52	 & 2.38 $\pm$ 0.15 $\pm$ 0.26	 \\ 
&&$\pi/2$	 & 18.38 $\pm$ 0.2 $\pm$ 0.67	 & 26.93 $\pm$ 0.31 $\pm$ 2.02	 & 24.28 $\pm$ 0.31 $\pm$ 1.38	 & 0.59 $\pm$ 0.16 $\pm$ 0.34	 \\ 
&&$3\pi/4$	 & 19.16 $\pm$ 0.2 $\pm$ 0.35	 & 23.96 $\pm$ 0.27 $\pm$ 1.78	 & 24.35 $\pm$ 0.3 $\pm$ 1.3	 & -2.01 $\pm$ 0.15 $\pm$ 0.31	 \\ 
\\
&0.4--0.5  &$0$	 & 17.36 $\pm$ 0.18 $\pm$ 0.5	 & 16.61 $\pm$ 0.2 $\pm$ 1.15	 & 16.81 $\pm$ 0.21 $\pm$ 1.31	 & 0.06 $\pm$ 0.12 $\pm$ 0.2	 \\ 
&&$\pi/4$	 & 17.01 $\pm$ 0.19 $\pm$ 0.42	 & 18.58 $\pm$ 0.23 $\pm$ 1.41	 & 17.42 $\pm$ 0.23 $\pm$ 1.59	 & 2.11 $\pm$ 0.13 $\pm$ 0.18	 \\ 
&&$\pi/2$	 & 15.76 $\pm$ 0.19 $\pm$ 0.32	 & 20.46 $\pm$ 0.27 $\pm$ 1.55	 & 17.73 $\pm$ 0.25 $\pm$ 1.16	 & 0.31 $\pm$ 0.14 $\pm$ 0.19	 \\ 
&&$3\pi/4$	 & 16.55 $\pm$ 0.19 $\pm$ 0.23	 & 18.51 $\pm$ 0.23 $\pm$ 1.25	 & 17.26 $\pm$ 0.23 $\pm$ 1.29	 & -1.79 $\pm$ 0.13 $\pm$ 0.17	 \\ 
\\
&0.5--0.6 &$0$	 & 15.78 $\pm$ 0.19 $\pm$ 0.45	 & 13.22 $\pm$ 0.19 $\pm$ 1.05	 & 13.85 $\pm$ 0.21 $\pm$ 0.68	 & 0.1 $\pm$ 0.12 $\pm$ 0.25	 \\ 
&&$\pi/4$	 & 15.32 $\pm$ 0.2 $\pm$ 0.42	 & 15.34 $\pm$ 0.23 $\pm$ 1.42	 & 14.25 $\pm$ 0.23 $\pm$ 0.88	 & 1.82 $\pm$ 0.13 $\pm$ 0.28	 \\ 
&&$\pi/2$	 & 13.75 $\pm$ 0.2 $\pm$ 0.39	 & 17.22 $\pm$ 0.28 $\pm$ 1.31	 & 14.18 $\pm$ 0.25 $\pm$ 0.74	 & 0.16 $\pm$ 0.14 $\pm$ 0.25	 \\ 
&&$3\pi/4$	 & 14.84 $\pm$ 0.19 $\pm$ 0.43	 & 14.32 $\pm$ 0.22 $\pm$ 1.4	 & 13.85 $\pm$ 0.22 $\pm$ 0.72	 & -1.36 $\pm$ 0.12 $\pm$ 0.24	 \\ 
\\
&0.6--0.8 &$0$	 & 13.87 $\pm$ 0.18 $\pm$ 0.4	 & 10.11 $\pm$ 0.16 $\pm$ 1.04	 & 9.81 $\pm$ 0.16 $\pm$ 0.49	 & 0.04 $\pm$ 0.1 $\pm$ 0.1	 \\ 
&&$\pi/4$	 & 12.33 $\pm$ 0.17 $\pm$ 0.35	 & 11.16 $\pm$ 0.18 $\pm$ 0.94	 & 9.88 $\pm$ 0.17 $\pm$ 1.06	 & 1.44 $\pm$ 0.11 $\pm$ 0.26	 \\ 
&&$\pi/2$	 & 11.28 $\pm$ 0.17 $\pm$ 0.2	 & 12.01 $\pm$ 0.21 $\pm$ 1.26	 & 10.51 $\pm$ 0.19 $\pm$ 1.24	 & 0.18 $\pm$ 0.11 $\pm$ 0.2	 \\ 
&&$3\pi/4$	 & 12.35 $\pm$ 0.17 $\pm$ 0.28	 & 10.59 $\pm$ 0.18 $\pm$ 0.78	 & 9.99 $\pm$ 0.17 $\pm$ 0.61	 & -1.23 $\pm$ 0.1 $\pm$ 0.17	 \\ 
\\
&0.8--1.5 &$0$	 & 11.04 $\pm$ 0.24 $\pm$ 0.36	 & 6.85 $\pm$ 0.19 $\pm$ 0.88	 & 6.37 $\pm$ 0.17 $\pm$ 0.74	 & -0.07 $\pm$ 0.12 $\pm$ 0.19	 \\ 
&&$\pi/4$	 & 9.83 $\pm$ 0.23 $\pm$ 0.31	 & 7.57 $\pm$ 0.23 $\pm$ 0.68	 & 6.05 $\pm$ 0.18 $\pm$ 0.73	 & 1.19 $\pm$ 0.13 $\pm$ 0.15	 \\ 
&&$\pi/2$	 & 7.92 $\pm$ 0.21 $\pm$ 0.52	 & 8.61 $\pm$ 0.28 $\pm$ 0.74	 & 5.64 $\pm$ 0.19 $\pm$ 0.44	 & 0.06 $\pm$ 0.13 $\pm$ 0.23	 \\ 
&&$3\pi/4$	 & 9.74 $\pm$ 0.25 $\pm$ 0.37	 & 7.04 $\pm$ 0.21 $\pm$ 0.78	 & 6.2 $\pm$ 0.18 $\pm$ 0.48	 & -1.28 $\pm$ 0.13 $\pm$ 0.18	 \\ 
\\
20\%--60\% & 0.2--0.3 &$0$	 & 16.24 $\pm$ 0.19 $\pm$ 0.29	 & 14.53 $\pm$ 0.21 $\pm$ 0.84	 & 22.44 $\pm$ 0.32 $\pm$ 0.97	 & 0.27 $\pm$ 0.13 $\pm$ 0.25	 \\ 
&&$\pi/4$	 & 15.18 $\pm$ 0.19 $\pm$ 0.28	 & 16.97 $\pm$ 0.24 $\pm$ 0.85	 & 22.64 $\pm$ 0.33 $\pm$ 0.7	 & 2.3 $\pm$ 0.14 $\pm$ 0.3	 \\ 
&&$\pi/2$	 & 12.81 $\pm$ 0.17 $\pm$ 0.24	 & 18.97 $\pm$ 0.28 $\pm$ 0.88	 & 21.85 $\pm$ 0.34 $\pm$ 0.76	 & 0.3 $\pm$ 0.14 $\pm$ 0.34	 \\ 
&&$3\pi/4$	 & 14.79 $\pm$ 0.18 $\pm$ 0.33	 & 16.34 $\pm$ 0.24 $\pm$ 0.95	 & 22.47 $\pm$ 0.33 $\pm$ 0.52	 & -1.8 $\pm$ 0.13 $\pm$ 0.35	 \\ 
\\
&0.3--0.4  &$0$	 & 13.98 $\pm$ 0.15 $\pm$ 0.27	 & 12.92 $\pm$ 0.14 $\pm$ 0.62	 & 15.9 $\pm$ 0.2 $\pm$ 0.59	 & 0.09 $\pm$ 0.09 $\pm$ 0.15	 \\ 
&&$\pi/4$	 & 13.15 $\pm$ 0.15 $\pm$ 0.22	 & 15.38 $\pm$ 0.18 $\pm$ 0.79	 & 15.77 $\pm$ 0.21 $\pm$ 0.46	 & 2.82 $\pm$ 0.1 $\pm$ 0.12	 \\ 
&&$\pi/2$	 & 11.04 $\pm$ 0.13 $\pm$ 0.26	 & 17.49 $\pm$ 0.22 $\pm$ 0.74	 & 15.11 $\pm$ 0.21 $\pm$ 0.52	 & 0.31 $\pm$ 0.11 $\pm$ 0.15	 \\ 
&&$3\pi/4$	 & 12.56 $\pm$ 0.14 $\pm$ 0.18	 & 15.03 $\pm$ 0.17 $\pm$ 0.74	 & 15.19 $\pm$ 0.2 $\pm$ 0.63	 & -2.4 $\pm$ 0.1 $\pm$ 0.16	 \\ 
\\
&0.4--0.5 &$0$	 & 12.58 $\pm$ 0.14 $\pm$ 0.17	 & 10.22 $\pm$ 0.12 $\pm$ 0.58	 & 11.31 $\pm$ 0.15 $\pm$ 0.45	 & 0.17 $\pm$ 0.08 $\pm$ 0.18	 \\ 
&&$\pi/4$	 & 11.55 $\pm$ 0.14 $\pm$ 0.22	 & 12.12 $\pm$ 0.15 $\pm$ 0.64	 & 11.31 $\pm$ 0.16 $\pm$ 0.51	 & 2.49 $\pm$ 0.09 $\pm$ 0.14	 \\ 
&&$\pi/2$	 & 9.32 $\pm$ 0.13 $\pm$ 0.28	 & 14.99 $\pm$ 0.21 $\pm$ 0.82	 & 11.39 $\pm$ 0.18 $\pm$ 0.63	 & 0.17 $\pm$ 0.1 $\pm$ 0.13	 \\ 
&&$3\pi/4$	 & 11.05 $\pm$ 0.13 $\pm$ 0.26	 & 12.25 $\pm$ 0.16 $\pm$ 0.58	 & 11.72 $\pm$ 0.17 $\pm$ 0.59	 & -2.04 $\pm$ 0.09 $\pm$ 0.19	 \\ 
\\
&0.5--0.6 &$0$	 & 11.28 $\pm$ 0.14 $\pm$ 0.17	 & 8.22 $\pm$ 0.12 $\pm$ 0.47	 & 9.08 $\pm$ 0.14 $\pm$ 0.29	 & 0.09 $\pm$ 0.09 $\pm$ 0.1	 \\ 
&&$\pi/4$	 & 10.15 $\pm$ 0.14 $\pm$ 0.29	 & 9.83 $\pm$ 0.15 $\pm$ 0.48	 & 9.38 $\pm$ 0.16 $\pm$ 0.39	 & 2.21 $\pm$ 0.09 $\pm$ 0.18	 \\ 
&&$\pi/2$	 & 8.12 $\pm$ 0.14 $\pm$ 0.27	 & 12.04 $\pm$ 0.21 $\pm$ 0.56	 & 8.96 $\pm$ 0.18 $\pm$ 0.49	 & -0.03 $\pm$ 0.1 $\pm$ 0.16	 \\ 
&&$3\pi/4$	 & 10.09 $\pm$ 0.14 $\pm$ 0.24	 & 10.2 $\pm$ 0.16 $\pm$ 0.64	 & 9.07 $\pm$ 0.16 $\pm$ 0.41	 & -2.05 $\pm$ 0.09 $\pm$ 0.13	 \\ 
\\
&0.6--0.8 &$0$	 & 9.57 $\pm$ 0.12 $\pm$ 0.21	 & 6.08 $\pm$ 0.09 $\pm$ 0.4	 & 6.53 $\pm$ 0.1 $\pm$ 0.31	 & 0 $\pm$ 0.06 $\pm$ 0.06	 \\ 
&&$\pi/4$	 & 8.55 $\pm$ 0.13 $\pm$ 0.12	 & 7.68 $\pm$ 0.13 $\pm$ 0.41	 & 6.46 $\pm$ 0.12 $\pm$ 0.25	 & 1.82 $\pm$ 0.07 $\pm$ 0.12	 \\ 
&&$\pi/2$	 & 6.71 $\pm$ 0.12 $\pm$ 0.31	 & 9.47 $\pm$ 0.18 $\pm$ 0.47	 & 6.61 $\pm$ 0.14 $\pm$ 0.27	 & 0.16 $\pm$ 0.08 $\pm$ 0.14	 \\ 
&&$3\pi/4$	 & 8.62 $\pm$ 0.13 $\pm$ 0.26	 & 7.56 $\pm$ 0.13 $\pm$ 0.45	 & 6.57 $\pm$ 0.12 $\pm$ 0.44	 & -1.82 $\pm$ 0.07 $\pm$ 0.16	 \\ 
\\
&0.8--1.5 &$0$	 & 7.81 $\pm$ 0.16 $\pm$ 0.27	 & 4.2 $\pm$ 0.1 $\pm$ 0.29	 & 4 $\pm$ 0.1 $\pm$ 0.2	 & 0 $\pm$ 0.07 $\pm$ 0.1	 \\ 
&&$\pi/4$	 & 7.18 $\pm$ 0.18 $\pm$ 0.33	 & 5.06 $\pm$ 0.15 $\pm$ 0.27	 & 4.3 $\pm$ 0.13 $\pm$ 0.24	 & 1.47 $\pm$ 0.09 $\pm$ 0.12	 \\ 
&&$\pi/2$	 & 4.74 $\pm$ 0.16 $\pm$ 0.22	 & 6.72 $\pm$ 0.26 $\pm$ 0.68	 & 3.84 $\pm$ 0.15 $\pm$ 0.65	 & 0.34 $\pm$ 0.1 $\pm$ 0.2	 \\ 
&&$3\pi/4$	 & 6.58 $\pm$ 0.17 $\pm$ 0.36	 & 5.36 $\pm$ 0.16 $\pm$ 0.25	 & 3.94 $\pm$ 0.12 $\pm$ 0.45	 & -1.32 $\pm$ 0.09 $\pm$ 0.09	 \\ 
 \end{tabular}\end{ruledtabular}
\label{data7}
\end{table*}

\begin{table*}[ht]
\caption{Azimuthal angle dependence of HBT radii of charged kaons, shown 
as value $\pm$ statistical uncertainty [absolute value] $\pm$ systematic 
uncertainty [\%] for the centrality bins shown in Fig.~\protect\ref{fig8}.}
\begin{ruledtabular}\begin{tabular}{cccccc}
Centrality & $\phi-\Psi_2$  & $\Rs^2$  &        $\Ro^2$  & $\Rl^2$  & $\Ros^2$ \\
           & [rad]          & [fm$^2$] &        [fm$^2$] & [fm$^2$] & [fm$^2$] \\
\hline
0\%--20\% & $0   	$ & 11.86 $\pm$ 0.68 $\pm$ 0.42	 & 10.77 $\pm$ 0.79 $\pm$ 1.14	 &10.43 $\pm$ 0.72 $\pm$ 1.34	 & 1.22 $\pm$ 0.58 $\pm$ 0.77	\\
& $\pi/4	$ & 10.02 $\pm$ 0.57 $\pm$ 1.04	 & 11.89 $\pm$ 0.82 $\pm$ 1.22	 & 8.74 $\pm$ 0.61 $\pm$ 1.70	 & 2.43 $\pm$ 0.52 $\pm$ 0.65	\\
& $\pi/2	$ &  7.98 $\pm$ 0.47 $\pm$ 0.92	 & 13.29 $\pm$ 0.97 $\pm$ 1.96	 & 9.25 $\pm$ 0.64 $\pm$ 1.47	 &-1.31 $\pm$ 0.52 $\pm$ 0.42	\\
& $3\pi/4	$ & 10.67 $\pm$ 0.61 $\pm$ 0.66	 & 12.45 $\pm$ 0.90 $\pm$ 0.86	 & 9.44 $\pm$ 0.65 $\pm$ 1.09	 &-2.15 $\pm$ 0.61 $\pm$ -0.69	\\ 
\\
20\%--60\% & $0   	$ & 10   $\pm$ 0.58 $\pm$ 0.69	 & 7.49 $\pm$ 0.54 $\pm$ 0.69	 & 5.71 $\pm$ 0.40 $\pm$ 0.6	 & 0.28 $\pm$ 0.44 $\pm$ 0.22	\\
& $\pi/4	$ & 7.52 $\pm$ 0.46 $\pm$ 0.56	 & 8.03 $\pm$ 0.58 $\pm$ 0.69	 & 6.07 $\pm$ 0.43 $\pm$ 0.71	 & 2.05 $\pm$ 0.40 $\pm$ 0.29	\\
& $\pi/2	$ & 4.69 $\pm$ 0.31 $\pm$ 0.68	 & 8.56 $\pm$ 0.59 $\pm$ 0.99	 & 5.61 $\pm$ 0.42 $\pm$ 0.81	 & 0.41 $\pm$ 0.34 $\pm$ 0.25	\\
& $3\pi/4	$ & 8.03 $\pm$ 0.49 $\pm$ 0.65	 & 8.51 $\pm$ 0.62 $\pm$ 0.77	 & 5.27 $\pm$ 0.40 $\pm$ 0.83	 &-1.52 $\pm$ 0.43 $\pm$ -0.24  \\ 
 \end{tabular}\end{ruledtabular}
\label{data6}
\end{table*}

\begin{table*}[ht]
\caption{Oscillation amplitudes relative to the event plane for charged pions, 
shown as value $\pm$ statistical uncertainty [absolute value] 
$\pm$ systematic uncertainty [\%] for the 0\%--20\% and 
20\%--60\% centrality bins.}
\begin{ruledtabular}\begin{tabular}{cccccc}
Centrality & \mt        & $\Rs^2$  &	$\Ro^2$  & $\Rl^2$  & $\Ros^2$ \\ 
           & [GeV/$c$]  & [fm$^2$] &	[fm$^2$] & [fm$^2$] & [fm$^2$] \\ 
\hline
0\%--20\% &0.30  &	 0.026 $\pm$ 0.007 $\pm$ 0.015 &	 0.095 $\pm$ 0.008 $\pm$ 0.024 &	 0.114 $\pm$ 0.010 $\pm$ 0.029 &	 0.099 $\pm$ 0.007 $\pm$ 0.009 \\ 
&0.38  &	 0.054 $\pm$ 0.006 $\pm$ 0.032 &	 0.101 $\pm$ 0.007 $\pm$ 0.018 &	 0.125 $\pm$ 0.008 $\pm$ 0.025 &	 0.113 $\pm$ 0.005 $\pm$ 0.013 \\ 
&0.47  &	 0.046 $\pm$ 0.007 $\pm$ 0.018 &	 0.104 $\pm$ 0.008 $\pm$ 0.025 &	 0.116 $\pm$ 0.008 $\pm$ 0.030 &	 0.117 $\pm$ 0.006 $\pm$ 0.010 \\ 
&0.57  &	 0.065 $\pm$ 0.008 $\pm$ 0.027 &	 0.125 $\pm$ 0.009 $\pm$ 0.020 &	 0.126 $\pm$ 0.009 $\pm$ 0.024 &	 0.107 $\pm$ 0.006 $\pm$ 0.016 \\ 
&0.70  &	 0.106 $\pm$ 0.008 $\pm$ 0.015 &	 0.082 $\pm$ 0.010 $\pm$ 0.032 &	 0.072 $\pm$ 0.009 $\pm$ 0.028 &	 0.107 $\pm$ 0.006 $\pm$ 0.016 \\ 
&0.93  &	 0.160 $\pm$ 0.014 $\pm$ 0.040 &	 0.098 $\pm$ 0.019 $\pm$ 0.040 &	 0.077 $\pm$ 0.015 $\pm$ 0.022 &	 0.129 $\pm$ 0.010 $\pm$ 0.013 \\ 
\\
20\%--60\% &0.30  &	 0.114 $\pm$ 0.007 $\pm$ 0.009 &	 0.131 $\pm$ 0.009 $\pm$ 0.024 &	 0.149 $\pm$ 0.010 $\pm$ 0.027 &	 0.140 $\pm$ 0.007 $\pm$ 0.022 \\ 
&0.38  &	 0.113 $\pm$ 0.007 $\pm$ 0.019 &	 0.150 $\pm$ 0.007 $\pm$ 0.015 &	 0.181 $\pm$ 0.008 $\pm$ 0.020 &	 0.207 $\pm$ 0.006 $\pm$ 0.010 \\ 
&0.47  &	 0.144 $\pm$ 0.007 $\pm$ 0.017 &	 0.180 $\pm$ 0.008 $\pm$ 0.015 &	 0.201 $\pm$ 0.009 $\pm$ 0.019 &	 0.205 $\pm$ 0.006 $\pm$ 0.012 \\ 
&0.57  &	 0.155 $\pm$ 0.008 $\pm$ 0.022 &	 0.185 $\pm$ 0.010 $\pm$ 0.011 &	 0.189 $\pm$ 0.010 $\pm$ 0.011 &	 0.216 $\pm$ 0.007 $\pm$ 0.014 \\ 
&0.70  &	 0.165 $\pm$ 0.009 $\pm$ 0.029 &	 0.212 $\pm$ 0.010 $\pm$ 0.021 &	 0.196 $\pm$ 0.010 $\pm$ 0.018 &	 0.220 $\pm$ 0.007 $\pm$ 0.017 \\ 
&0.93  &	 0.226 $\pm$ 0.015 $\pm$ 0.026 &	 0.211 $\pm$ 0.019 $\pm$ 0.027 &	 0.172 $\pm$ 0.016 $\pm$ 0.021 &	 0.216 $\pm$ 0.010 $\pm$ 0.012 \\ 
 \end{tabular}\end{ruledtabular}
\label{data8}
\end{table*}

\begin{table*}[ht]
\caption{Oscillation amplitudes relative to the event plane for charged kaons, 
$\pm$ systematic uncertainty [\%] for the 0\%--20\% and
20\%--60\% centrality bins shown in Fig.~\protect\ref{fig12}}. 
\begin{ruledtabular}\begin{tabular}{cccccc}
Centrality & \mt        & $\Rs^2$  &	$\Ro^2$  & $\Rl^2$  & $\Ros^2$ \\ 
           & [GeV/$c$]  & [fm$^2$] &	[fm$^2$] & [fm$^2$] & [fm$^2$] \\ 
\hline
 0\%--20\% & 0.91  & 0.193 $\pm$ 0.034 $\pm$ 0.078 & 0.106 $\pm$ 0.044 $\pm$ 0.033 & 0.128 $\pm$ 0.052 $\pm$ 0.044 & 0.227 $\pm$ 0.040 $\pm$ 0.067 \\ 
\\
20\%--60\% & 0.91  & 0.360 $\pm$ 0.035 $\pm$ 0.080 & 0.070 $\pm$ 0.042 $\pm$ 0.035 & 0.076 $\pm$ 0.045 $\pm$ 0.038 & 0.238 $\pm$ 0.040 $\pm$ 0.030 \\ 
 \end{tabular}\end{ruledtabular}
\label{data9}
\end{table*}

\clearpage


%
 
\end{document}